\documentclass[aps,prb,superscriptaddress,citeautoscript,twocolumn,notitlepage,longbibliography]{revtex4-2}

\usepackage{graphicx}
\usepackage{dcolumn}
\usepackage{bm}

\usepackage{graphicx}
\usepackage{amsmath}
\usepackage{amssymb}
\usepackage[normalem]{ulem}
\usepackage{braket}
\usepackage{color}
\usepackage{gensymb}
\usepackage[colorlinks=True,linkcolor=red,citecolor=blue,urlcolor=blue]{hyperref}

\usepackage{makecell}
\usepackage{dsfont}
\usepackage{mathtools,esint}
\newcommand{\bea}{\begin{eqnarray*}}
\newcommand{\eea}{\end{eqnarray*}}
\newcommand{\bne}{\begin{equation*}}
\newcommand{\ede}{\end{equation*}}

\newcommand{\bnen}{\begin{equation}}
\newcommand{\eden}{\end{equation}}
\newcommand{\bean}{\begin{eqnarray}}
\newcommand{\eean}{\end{eqnarray}}
\newcommand{\bsen}{\begin{subequations}}
\newcommand{\esen}{\end{subequations}}

\newcommand{\bna}{\begin{array}}
\newcommand{\eda}{\end{array}}
\newcommand{\bnm}{\begin{enumerate}}
\newcommand{\edm}{\end{enumerate}}
\newcommand{\bni}{\begin{itemize}}
\newcommand{\edi}{\end{itemize}}

\renewcommand{\vec}[1]{\text{\boldmath{$ #1 $}}}

\definecolor{darkgreen}{rgb}{0.0, 0.5, 0.0}

\newcommand*{\vertbar}{\rule[-1ex]{0.5pt}{2.5ex}}
\newcommand*{\horzbar}{\rule[.5ex]{2.5ex}{0.5pt}}

\DeclareMathAlphabet\mathbfcal{OMS}{cmsy}{b}{n}

\begin{document}


\title{Topological charge distributions of an interacting two-spin system}

\author{György Frank}

\affiliation{Department of Theoretical Physics,
Budapest University of Technology and Economics, 
Hungary}

\affiliation{MTA-BME Exotic Quantum Phases 
 Group,
 Budapest University of Technology and Economics, Hungary}

\author{D\'aniel Varjas}
\affiliation{QuTech and Kavli  Institute  of  Nanoscience,  Delft  University  of  Technology, 2600  GA  Delft,  The  Netherlands}
\affiliation{Department of Physics, Stockholm University, AlbaNova University Center, 106 91 Stockholm, Sweden}

\author{P\'eter Vrana}

\affiliation{Institute of Mathematics, Budapest University of Technology and Economics, Hungary}

\author{Gerg\H{o} Pint\'er}

\affiliation{Department of Theoretical Physics,
Budapest University of Technology and Economics, 
Hungary}

\affiliation{Institute of Mathematics, Eötvös Loránd University, 
Pázmány Péter sétány 1/c, H-1117 Budapest, Hungary}

\author{Andr\'as P\'alyi}

\affiliation{Department of Theoretical Physics,
Budapest University of Technology and Economics, 
Hungary}

\affiliation{MTA-BME Lendület Topology and Correlation Research Group,
Budapest University of Technology and Economics, 1521 Budapest, Hungary}

\date{\today}

\begin{abstract}
Quantum systems are often described by 
parameter-dependent Hamiltonians.
Points in parameter space where two levels are 
degenerate can carry a topological charge. 
Here we theoretically study an interacting
two-spin system where the degeneracy points
form a nodal loop 
or a nodal surface in the 
magnetic parameter space, similarly
to such structures discovered in the band structure
of topological semimetals. 
We determine the topological charge distribution
along these degeneracy geometries.
We show that these non-point-like degeneracy patterns 
can be obtained not only by fine-tuning, but they
can be stabilized by spatial symmetries.
Since simple spin systems such as the one
studied here are ubiquitous in condensed-matter
setups, we expect that our findings, and 
the physical consequences of these nontrivial degeneracy geometries,
are testable in experiments with quantum dots,
molecular magnets, and adatoms on metallic surfaces.
\end{abstract}

\maketitle

\section{Introduction}
Quantum systems are often described
by parameter-dependent Hamiltonians,
with many models incorporating
multiple tunable parameters 
\cite{Herring,Berry,HasanRMP2010,Armitage,Asboth,Riwar,Scherubl}. 
For example, the three Cartesian
components of the external magnetic field provide
$N=3$ parameters in the Hamiltonian 
of an interacting multi-spin system
\cite{Wernsdorfer,Bruno,Gritsev,Scherubl,Frank}.

Let us summarize a few \emph{generic} features
for the case when the only constraint on the
parametrized Hamiltonian is its Hermiticity. 
In this case, it requires at least 
$N=3$ parameters to find points
in the parameter space where two 
of the energy levels are 
degenerate~\cite{Neumann,ArnoldSelMath1995}.
If the dimension of the parameter space
is exactly $N=3$, then the
\emph{generic} degeneracy points are isolated.
If $N>3$, then the degeneracy
points form $(N-3)$-dimensional
geometrical patterns in the $N$-dimensional
parameter space, e.g., 
lines in a 4-dimensional parameter space, 
surfaces in a 5-dimensional parameter space,
etc. 
Moreover, in the vicinity of a generic isolated degeneracy point
in a three-dimensional parameter space
(a so called \emph{Weyl point}),
the energy splitting between the two levels depends
linearly on the distance from the Weyl point.

We may associate a topological
charge to a point-like degeneracy
in a three-dimensional parameter space 
\cite{SimonPRL1983,Bruno,Scherubl,Frank}. 
For example, take a single localized electron
in a magnetic (Zeeman) field, 
\begin{equation}
\label{eq:spin}
    H = \vec B \cdot \vec S,
\end{equation}
where $\vec B$ is the magnetic field and $\vec S$ is the spin--$1/2$ vector operator,
that is, $1/2$ times the Pauli matrices.
In this example, the degeneracy point
is at the origin, $\vec B_0 =0$.
Calculating the surface integral of the
ground-state Berry curvature vector field
on a closed surface surrounding this degeneracy
point yields 1, independent of the shape of the surface.
For a closed surface whose interior
does not contain the degeneracy
point, this integral is zero. 
The relations of these observations
to topology, and 
to the electrostatics of a point charge,
justify the terminology that
the degeneracy point carries 
unit topological charge. 
Note also that the Hamiltonian in Eq.~\eqref{eq:spin}
exemplifies the above-mentioned
generic feature of linear energy splitting.

Fine-tuning or symmetries can lead
to anomalous, 
\emph{non-generic} situations when degeneracy points in 
a three-dimensional parameter space ($N=3$) are
(i)  isolated, but the energy splitting is not linear, 
but of higher order 
\cite{ChenFang_multiweyl,ZhongboYan,Ahn,ZeMinHuang}, or 
(ii)  not isolated, but they 
form a continuous line or surface
\cite{BeriPRB2010,CarterPRB2012,ChenFang,WeikangWu,ChenFangChinese2016,bzduvsek2016nodal,
QiFengLiangPRB2016,ZhongboYan,YingMingXieArxiv2020}. 
These anomalous features have been demonstrated
in electronic band structure models of
three-dimensional solids, where the
parameters are the Cartesian components of 
crystal momentum,
and also in interacting spin systems 
with a three-dimensional magnetic parameter space~\cite{Frank}. 

\begin{figure}[tbh]
	\begin{center}
		\includegraphics[width=0.9\columnwidth]{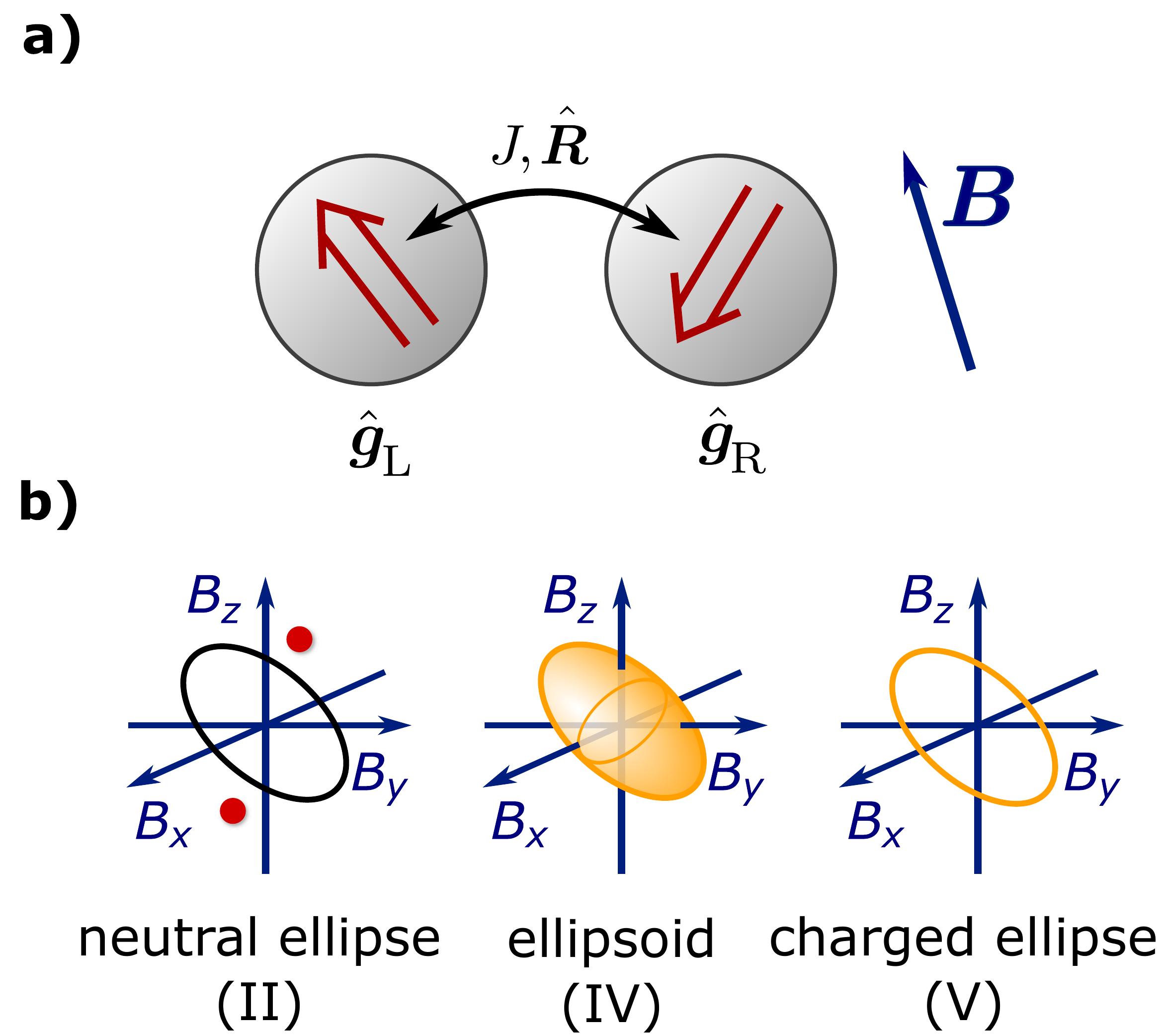}
	\end{center}
	\caption{{\bf Interacting two-spin system 
	and its non-point-like magnetic degeneracy patterns.} 
	{\bf a)} Two localized spinful electrons
	in the presence of spin-orbit coupling, interacting
	with each other via exchange interaction, and
	an external magnetic (Zeeman) field $\vec B$.
	{\bf b)}
	Non-point-like magnetic degeneracy patterns
	of the two-spin system.
	(II) Two degeneracy points with unit ($+1$)
	topological charge each
	(red) and a neutral degeneracy ellipse.
	(IV) Degeneracy ellipsoid surface carrying a total 
	topological charge of $+2$.
	(V) Degeneracy ellipse carrying a  total
	topological charge of $+2$.
	Classes (II), (IV), (V) were defined in 
	Table I of Ref.~\onlinecite{Frank}.
	\label{fig:geometries}}
\end{figure}

In this work, we consider 
the three-dimensional magnetic parameter space,
and focus on case (ii), i.e., when either a degeneracy
line or a degeneracy surface is present. 
As an experimental motivation and an
illustrative example, we take the
spin-orbit-coupled interacting two-spin 
problem we studied in the experiment of 
Ref.~\onlinecite{Scherubl} and in the theory work 
Ref.~\onlinecite{Frank}.
The setup is illustrated in Fig.~\ref{fig:geometries}~a.
In that two-spin problem, 
a magnetic degeneracy line appears
in cases (II) and (V),
and a magnetic degeneracy surface
appears in case (IV), where the cases are
defined in  Table I of Ref.~\onlinecite{Frank}.
The degeneracy patterns are sketched in
Fig.~\ref{fig:geometries}~b.

The topological charge of an isolated
magnetic degeneracy point is concentrated
in that single point. 
However, for a degeneracy line
or degeneracy surface, charged or neutral, 
it is a natural question to ask:
how is the net topological charge 
\emph{distributed} on the line
or surface?
As the central results of this work, 
we provide answers for each of the
three cases. 

For the neutral ellipse in case (II),
we find that the topological 
charge distribution along the ellipse
is identically zero. 
For the charged ellipse in 
case (V),
the net topological charge 
$+2$ is evenly distributed between two
opposite points of the degeneracy
ellipse, and all 
further points of the ellipse
are neutral. 
For the charged ellipsoid in case (IV), 
the net topological charge of +2
is distributed continuously, 
in striking similarity to how electric charge
is distributed on the surface of 
a charged metallic ellipsoid. 
Furthermore, we show that these non-point-like
degeneracy geometries can be obtained not only
by fine-tuning, but they can be stabilized by spatial symmetries; 
we exemplify this for the case when the two-spin system
has a $C_{3v}$ symmetry.

The rest of the paper is structured as follows.
In section~\ref{sec:setup}, we define the Hamiltonian of the 
spin-orbit coupled interacting two-spin model that we consider throughout 
this work, and review our earlier results regarding the
possible geometrical patterns formed by the 
ground-state degeneracy points of the magnetic parameter space.
In what follows, we call the degeneracy patterns labelled
(II), (IV), and (V) of Ref.~\onlinecite{Frank} as the 
\emph{non-point-like}
degeneracy patterns of two-spin model. 
In sections~\ref{sec:linear} and~\ref{sec:ellipsoid},
we present and derive the topological charge distributions
characterizing these non-point-like degeneracy patterns.
We provide a discussion of our results in
\ref{sec:discussion}, and conclude in~\ref{sec:conclusions}.

\section{Setup: spin-orbit coupled two-spin system}
\label{sec:setup}

The system under investigation 
consists of two spinful electrons distributed in a double-well 
potential, interacting with each other, in the presence
of spin-orbit interaction 
and a homogeneous magnetic field.
We will describe this system with the following
dimensionless $4\times 4$ Hamiltonian matrix~\cite{Kavokin,Frank,Scherubl}:
\begin{eqnarray}
\label{eq:hamiltonian}
H &= \vec B \cdot \left(\hat{\vec g}_\text{L} \vec S_\text{L} + \hat{\vec g}_\text{R} \vec S_\text{R}\right)+ J \vec S_\text{L} \cdot \hat{\vec R} \vec S_\text{R}.
\end{eqnarray}
Here, the first term is the Zeeman interaction with the external 
homogeneous magnetic field $\vec B$, where $\vec S_\text{L}$ and $\vec S_\text{R}$ are the 
spin vector operators represented by $1/2$ times the spin-1/2 
Pauli matrices, and 
$\hat{\vec g}_\text{L}$ and $\hat{\vec g}_\text{R}$ are the real-valued $g$-tensors that are affected
by spin-orbit coupling.
The $g$-tensors are not necessarily symmetric, but 
we assume that both have a positive determinant.
The second term is the exchange interaction
between the two electrons, which deviates
from standard Heisenberg exchange due to 
spin-orbit interaction. 
In that term, 
$J>0$ is the strength of the exchange interaction, 
and $\hat{\vec R}$ is a  real, $3\times3$ special orthogonal matrix accounting for the spin-orbit
interaction in the exchange term.
The origin of this Hamiltonian is discussed 
in detail in~\cite{Kavokin,Scherubl,Frank}.
Note that numerous experiments have shown that $g$-tensors of
electrons confined in semiconductors can be tuned in situ by electric fields~\cite{Kato-gfactorresonance,VeldhorstPRB,Crippa,Schroer-gfactor,LilesArxiv2020}.

In what follows, we will refer to the magnetic-field space as
the \emph{parameter space}, and will use the term 
\emph{secondary parameters} for further parameters of the Hamiltonian:
$g$-tensors, exchange strength $J$, and the exchange rotation matrix
$\hat{\vec R}$.

Throughout this work, we will focus on the values of the
magnetic field $\vec B_0$ where the ground state of 
this $4 \times 4$ Hamiltonian is degenerate. 
Here we recall results from Ref.~\onlinecite{Frank} 
that identify such
magnetic degeneracy points. 
We have found that 
if it holds for a 
unit vector $\vec b$ that
\bean
\label{eq:parallelcondition}
\vec{b}^{\text{T}}\hat{\vec g}_{\text{R}}\hat{\vec R}^{-1}& \;||\; 
&\vec b^{\text{T}} 
\hat{\vec g}_{\text{L}},
\eean
then there is a unique ground-state degeneracy
point at a certain magnetic field 
$\vec B_+ = B_0 \vec b$
with $B_0>0$,
and another one at $\vec B_- = -B_0 \vec b$.
(Note that in our notation, $\parallel$ includes 
that 
the two vectors point to the same direction.)
In turn, condition \eqref{eq:parallelcondition} 
is fulfilled if and only if $\vec b$
is a left eigenvector of the matrix 
\bean
\label{eq:matrixm}
\hat{\vec M} = \hat{\vec g}_\text{L} \hat{\vec R} \hat{\vec g}_\text{R}^{-1}
\eean
corresponding to a positive eigenvalue 
$a$. The absolute value of the magnetic field where the ground-state degeneracy occurs is
\begin{eqnarray}\label{eq:bnorm}
B_0=\left(1+\frac{1}{a}\right)\frac{1}{2g_\text{R}},
\end{eqnarray}
where $g_{\text R}=|\hat{\vec g}_\text{R}^{\text T}\vec b|$ (See Appendix C of Ref.~\onlinecite{Frank}).

The above condition Eq.~\eqref{eq:parallelcondition}
is sufficient to guarantee the existence of two degeneracy points. 
We do not have a rigorous proof that Eq.~\eqref{eq:parallelcondition} 
is also a necessary condition, 
but an extensive numerical search for degeneracy points found no counterexample, 
so we conjecture that it is.

The matrix $\hat{\vec M}$ 
defined in Eq.~\eqref{eq:matrixm}
is a $3\times 3$ non-symmetric real-valued matrix with positive determinant. 
The possible degeneracy geometries
are classified by its eigenstructure, 
i.e., the Jordan normal form of this matrix, see Table I in Ref.~\onlinecite{Frank}.
As shown there, 
the degeneracy points
can be isolated, as in the electronic dispersion relation 
of a Weyl semimetal~\cite{Armitage}
or multi-Weyl semimetal~\cite{ChenFang_multiweyl}, 
or they can form lines or surfaces, as in nodal-loop~\cite{ChenFang} or nodal-surface~\cite{WeikangWu} semimetals.

\section{Linear charge density along degenerate lines}
\label{sec:linear}

How is the topological charge distributed
along a degeneracy line? 
To answer this question, we follow intuition from
classical electrostatics. 
Since in our two-spin problem the
degeneracy lines are closed loops, we 
take such an example from electrostatics. 

\begin{figure}[tbh]
	\begin{center}
		\includegraphics[width=0.8\columnwidth]{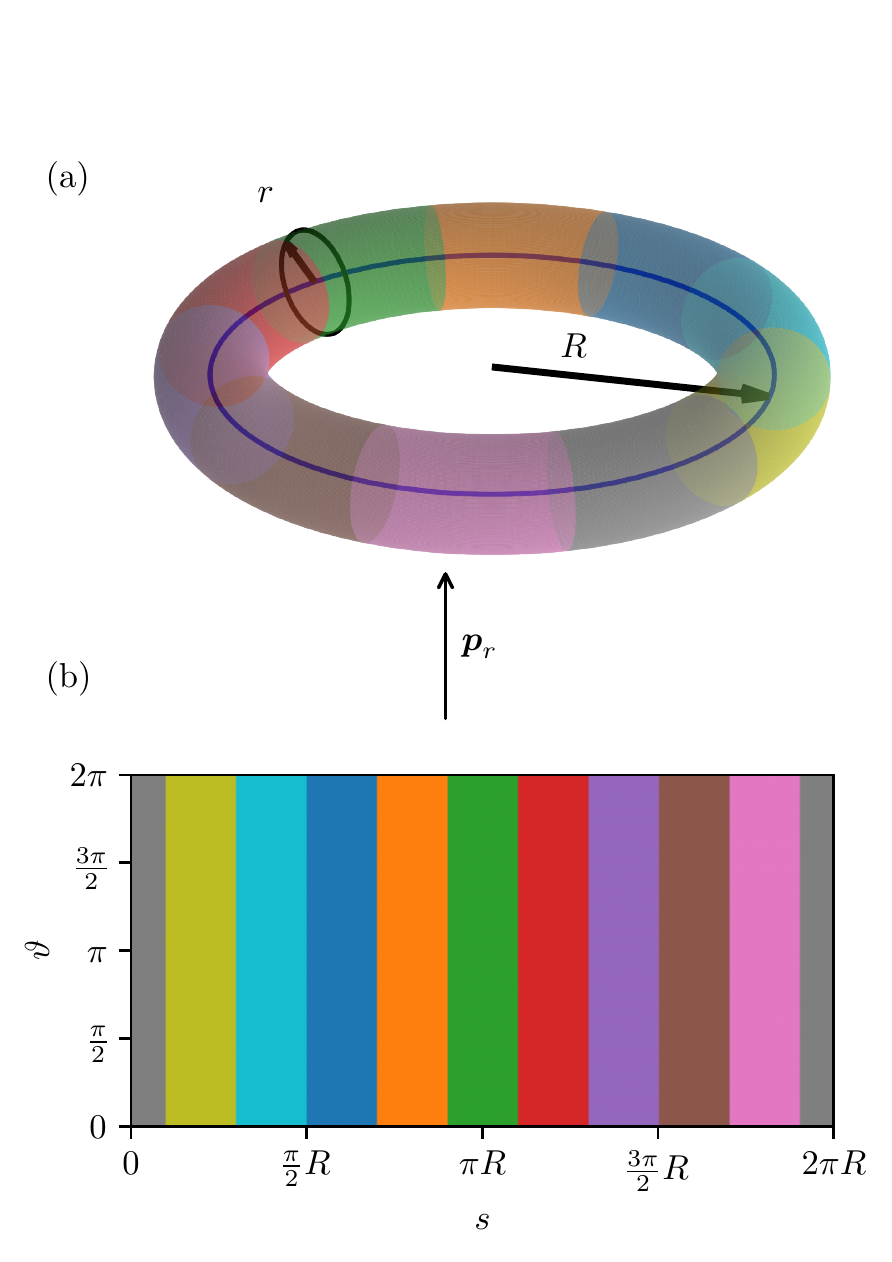}
	\end{center}
	\caption{{\bf Inferring the linear charge
	density of a charged loop from the electric
	field it creates.} 
	(a) Charged loop (blue) and 
	a torus (colored) enclosing the loop, with meridian radius $r$. 
	(b) The torus is parametrized
	by the map 
	$\vec p_r: [0,2\pi R) \times [0,2\pi) \to \mathbb{R}^3$, 
	using the longitude path length $s$ and 
	the meridian angle $\theta$.
	The electric flux density on a torus 
	reveals the linear charge density 
    of the loop as meridian radius
    of the torus shrinks to zero, 
    $r\to 0$, see Eq.~\eqref{eq:electrostatics}.
 	\label{fig:torus}}
\end{figure}

As shown in Fig.~\ref{fig:torus},
consider a loop $l$ (blue), 
chosen to be circular with radius 
$R$ for concreteness,
parametrized by the
path length variable $s \in [0,2\pi R)$.
Assume that this loop
has a linear electrostatic charge distribution 
$\nu(s)$.
The charge creates an electric field $\vec E(\vec r)$.
Can we deduce the linear charge density
if only the induced electric field is known?
Yes, in the following way:
\begin{eqnarray}
\label{eq:electrostatics}
\frac{\nu(s)}{\varepsilon_0} =
\lim_{r\to 0}
\int_0^{2\pi} d\vartheta \, 
\vec E(\vec p_r(s,\vartheta))
\cdot
\left(
\partial_\vartheta \vec p_r \times \partial_s \vec p_r
\right)_{s,\vartheta}.
\end{eqnarray}
In this formula, $\vec p_r:[0,2\pi R) \times [0,2\pi)$
is the parametrization of a torus surrounding
the loop as shown in Fig.~\ref{fig:torus},
with $s \in [0,2\pi R)$ used as the \emph{longitude path length}
and $\vartheta \in [0,2\pi)$ used as the
\emph{meridian angle} of the torus.
Furthermore, $r$ is the meridian radius characterizing the
thickness of the torus, that is, $r\to0$ corresponds
the thickness shrinking and the torus 
surrounding the loop infinitely tightly.
Note that the dimension of $s$ is length (meter)
whereas $\vartheta$ is an angle parameter hence is 
dimensionless.
We prove this relation between the linear charge density and electric field in classical electrostatics in Appendix~\ref{app:electrostatics}.

Using the relation of Eq.~\eqref{eq:electrostatics}
we identify the linear topological charge density
along a degeneracy line.
Patterns (II) and (V) are degeneracy
loops (ellipses), hence we can 
again surround any of them by a shrinking torus,
described by the parametrization 
$\vec p_r(s,\vartheta)$,
where $s$ has the dimension of magnetic field 
(Tesla) and $\vartheta$ is dimensionless. 
For an isolated degeneracy point, 
the ground-state topological charge 
or Chern number associated to the point
reads
\begin{equation}
\mathcal{Q} = \frac{1}{2\pi} \int_S 
d\vec A \cdot \vec{\mathcal{B}},
\end{equation}
where the integral is calculated for a closed surface $S$
enclosing the isolated degeneracy point, 
and $\vec{\mathcal B}$ is the Berry curvature
vector field associated to the
the ground-state wave function $\psi_0(\vec B)$ defined as
\begin{equation}
\label{eq:berrycurvature}
    \vec{\mathcal{B}}(\vec B) =  i
    \braket{\vec{\nabla}_{\vec{B}}\psi_0 (\vec B)| \times | \vec{\nabla_{\vec B}} \psi_0(\vec B)},
\end{equation}
or writing component-wise
\begin{equation}
\label{eq:componentwise}
    \mathcal{B}_i(\vec B) = i\epsilon_{ijk} \braket{\partial_{B_j}\psi_0|\partial_{B_k}\psi_0}.
\end{equation}

Therefore, for a degeneracy line, 
the formula revealing the linear 
topological charge density reads
\begin{eqnarray}
\label{eq:topologicalchargedensity}
\nu(s) = \frac{1}{2\pi}
\lim_{r\to 0} \int_0^{2\pi}
d \vartheta\,  \vec{\mathcal B}(\vec p_r (s,\vartheta))
\cdot 
\left(
\partial_\vartheta \vec p_r \times \partial_s \vec p_r
\right)_{s,\vartheta}.\;\;\;\;
\end{eqnarray}
This is the quantity that we study in the following.

For future reference, we introduce
the \emph{two-dimensional (2D) Berry curvature}
$\mathcal{B}_\text{2D}$ via
\begin{equation}
\label{eq:2dberrycurvature}
    \mathcal{B}_\text{2D}(s,\vartheta) =
    \vec{\mathcal B}(\vec p_r (s,\vartheta))
    \cdot 
    \left(
    \partial_\vartheta \vec p_r \times \partial_s \vec p_r
    \right)_{s,\vartheta},
\end{equation}
i.e., the integrand in Eq.~\eqref{eq:topologicalchargedensity},
and the \emph{apparent topological charge density}
$\tilde{\nu}_r(s)$, 
which is the right hand side of Eq.~\eqref{eq:topologicalchargedensity},
without taking the limit $r \to 0$:
\begin{eqnarray}
\label{eq:apparent}
    \tilde{\nu}_r(s) = 
    \frac{1}{2\pi} \int_0^{2\pi} d\vartheta
    \mathcal{B}_\text{2D}(s, \vartheta),
\end{eqnarray}
related to the charge density defined above as
\begin{equation}
\nu(s) = \lim_{r\to 0} \tilde{\nu}_r(s).
\end{equation}

Since we use the Hamiltonian
of Eq.~\eqref{eq:hamiltonian} depending on dimensionless parameters $\vec{B}$ as 
our starting point, 
all these newly introduced quantities
are also dimensionless.
Reinstating physical dimensions in 
Eq.~\eqref{eq:hamiltonian} is done
by multiplying the first term with the
Bohr magneton and reinterpreting
$\vec{B}$ as a magnetic field and $J$ as
an energy.
Then, the physical dimension of
the Berry curvature and the Berry flux
density is magnetic field$^{-2}$,
whereas the dimension of the 
2D Berry curvature, the apparent topological
charge density and the topological 
charge density is magnetic field$^{-1}$.

\subsection{Pattern (II): neutral ellipse}
\label{sec:neutral}

\begin{figure*}[tbh]
	\begin{center}
		\includegraphics[width=2\columnwidth]{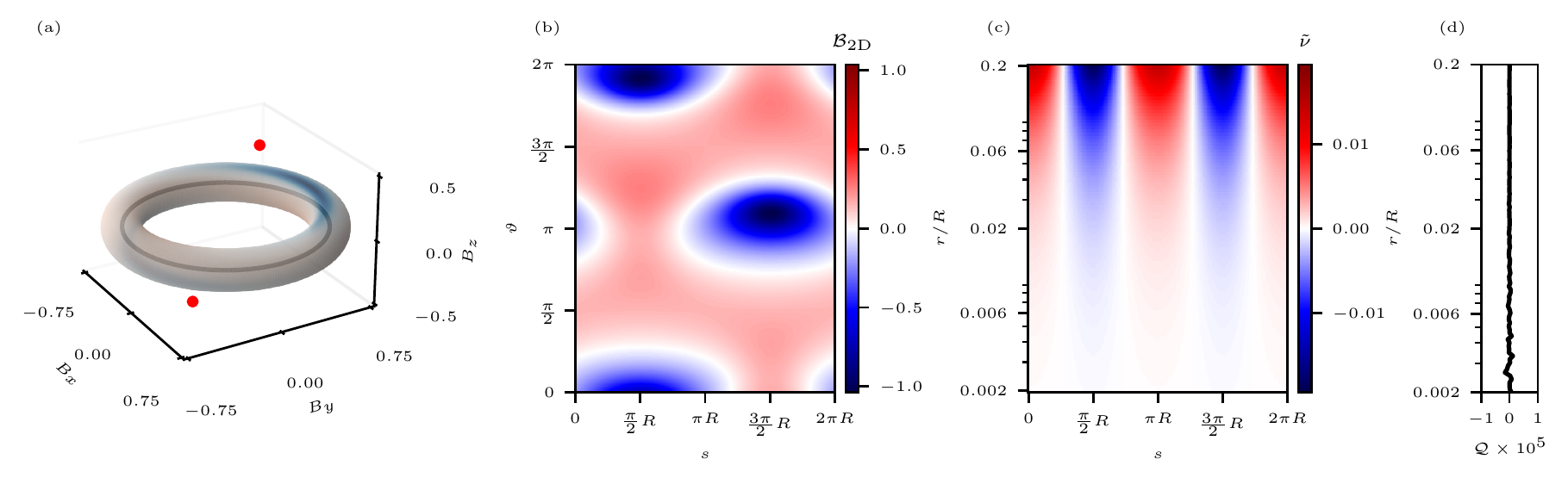}
	\end{center}
	\caption{{\bf Neutral degeneracy ellipse
	has vanishing linear topological charge distribution.} 
	(a) Berry flux density $\mathcal{B}_n$ (temperature map)
	on the surface of a torus
	surrounding a neutral degeneracy circle (black).
	Meridian radius, $r/R=0.2$
    (b) Two-dimensional Berry curvature
    $\mathcal{B}_\text{2D}(s,\vartheta)$
    [Eq.~\eqref{eq:2dberrycurvature}]
    on the pre-image of the torus, i.e., as a function 
    of longitude path length $s$ and meridian angle $\vartheta$. 
    (c) Apparent linear topological charge density $\tilde{\nu}_r(s)$
    [Eq.~\eqref{eq:apparent}],
    as function of longitude path length $s$ and
    meridian radius $r$.
    As $r\to 0$, this function converges to the
    constant zero function (white), showing 
    that the linear topological charge density
    vanishes, $\nu(s) = 0$.
    (d) A benchmark for the numerical integration:
    numerically evaluated ground-state
    Chern number $\mathcal{Q}$ on the torus, 
    as function of the meridian radius $r$.
    Numerical error grows slightly as $r\to 0$, due
    to the divergence of the Berry curvature in the vicinity
    of the degeneracy circle, but it remains well below $10^{-5}$
    even for the smallest $r$ values considered.
	\label{fig:neutral}}
\end{figure*}

First we consider the neutral ellipse degeneracy pattern (II)
in Fig.~\ref{fig:geometries}. 
For this pattern with zero total charge 
one can envision two qualitatively 
different scenarios: 
(a) The charge distribution
is identically zero at all
points of the ellipse. 
(b) There is a non-zero linear charge 
density along the ellipse,
but the negative and positive
contributions cancel each other
when added up for the entire ellipse.
Speculation based on classical 
electrostatics intuition actually 
suggest scenario (b): if we think 
of the ellipse as a globally charge-neutral
`metal', then the two point charges outside
the ellipse would `polarize' the ellipse.

In contrast, here we provide
evidence that scenario (a) is 
the case, the local charge distribution along
the degeneracy ellipse
vanishes. 
This conclusion will be drawn from 
Fig.~\ref{fig:neutral}~c, but let us 
arrive there through a few intermediate steps.

In Fig.~\ref{fig:neutral}~a, we show
the degeneracy patterns,
two red points and a black ellipse.
We use the specific choice of
parameters where the $g$-tensors are
\begin{eqnarray}
\label{eq:neutralgtensors}
\hat{\vec g}_{\text{L,II}}=
\begin{pmatrix}
2&0&0\\
0&2&0\\
0&1&4
\end{pmatrix},
\hspace{6mm}
\hat{\vec g}_{\text{R,II}}=\mathds{1}_{3\times 3},
\end{eqnarray}
and the exchange interaction
is characterized by 
$J_{\text{II}}=1$ and $\hat{\vec R}_{\text{II}}=\mathds{1}_{3\times 3}$.
For simplicity, energy and magnetic field are
dimensionless, unless noted otherwise. 

The total topological charge carried by 
the red degeneracy points in Fig.~\ref{fig:neutral}~a
is $+2$.
These degeneracy points are located
at opposite magnetic fields,
\begin{eqnarray}
\label{eq:weylpoints}
\vec B_\pm=\pm \frac{\sqrt{5}}{8}
\begin{pmatrix}
0\\
1\\
2
\end{pmatrix}.
\end{eqnarray}
and each of them carry a topological
charge $+1$.
The degeneracy ellipse
shown as the black loop in Fig.~\ref{fig:neutral}
is actually a circle in the $xy$ plane for
this parameter set, 
centered at the origin, 
with radius $R=\frac{3}{4}$.

Figure~\ref{fig:neutral}~a shows the Berry flux density
on a torus surrounding the degeneracy circle. 
The Berry flux density is defined as the normal-to-surface
component of the Berry curvature vector field.
For example, for a point $\vec B$ on the torus, 
the Berry flux density  reads 
\begin{eqnarray}
\label{eq:berryfluxdensity}
\mathcal B_n(\vec B)=\mathbfcal{B}(\vec B)\cdot\vec n(\vec B)
\, \,\, \,  (\vec B \in \text{torus}),
\end{eqnarray}
where $\vec n(\vec B)$ is the normal vector of the torus 
at point $\vec B$.
The torus in Fig.~\ref{fig:neutral}~a
is colored according to the nonzero Berry flux density.
(Numerical techniques to obtain Fig.~\ref{fig:neutral}
are described in Appendix~\ref{sec:numerics}.)

On the way toward the linear topological charge density, 
to be expressed via Eq.~\eqref{eq:topologicalchargedensity},
we specify the parametrization of the torus surrounding 
the degeneracy line as
\begin{equation}
\label{eq:parametrization}
    \vec p_r(s,\vartheta) = 
    R \vec{e}_\text{rad}(s) +
    r \left[ \cos \vartheta \, \vec{e}_z +\sin \vartheta \, \vec e_\text{rad}(s) \right],
\end{equation}
with
\begin{eqnarray}
    \vec e_\text{rad}(s) = \left(\begin{array}{c}
        \cos(s/R) \\ \sin(s/R) \\ 0
    \end{array} \right), \;\;\;\;\;
    \vec e_z = \left(\begin{array}{c}
        0 \\ 0 \\ 1
    \end{array} \right).
\end{eqnarray}
Note that the normal vector
of the torus can be expressed from the parametrization
via
\begin{equation}
\label{eq:normalvector}
   \vec n(\vec p_r(s,\vartheta)) = 
   \frac{
        \partial_\vartheta \vec p_r \times \partial_s \vec p_r
    }
    {
        \left|
        \partial_\vartheta \vec p_r \times \partial_s \vec p_r
        \right|
    }.
\end{equation}

With the parametrization in Eq.~\eqref{eq:parametrization}, 
in Fig.~\ref{fig:neutral}~b we plot 
the 2D Berry curvature $\mathcal{B}_\text{2D}$ 
(see Eq.~\eqref{eq:2dberrycurvature}) on the torus, 
with meridian radius $r = 0.2 R$.
The data in Fig.~\ref{fig:neutral}~b is used
to infer the linear topological charge density, by
numerically performing the integration over the parameter $\vartheta$
and dividing by $2\pi$ to obtain 
the apparent charge density $\tilde{\nu}_r(s)$, 
and then taking the limit $r\to 0$.
The apparent charge density as function of
$s$ and $r$ is shown in Fig.~\ref{fig:neutral}~c.
Although the value of the apparent charge density is nonzero for 
finite $r$, it does converge to zero for 
all values of $s$ as $r\to 0$.
This is numerical evidence that the degeneracy 
circle is charge neutral.
In section~\ref{sec:discussion} we provide
further analytical evidence to support this 
claim. 

To illustrate the accuracy of our result shown in Fig.~\ref{fig:neutral}~c,
we numerically evaluate the ground-state Chern number $\mathcal{Q}$ on the 
torus as the function of the meridian radius $r$, 
by integrating the apparent charge density over 
the longitude path length $s$.
The result, shown in Fig.~\ref{fig:neutral}~d,
is indeed zero, exhibiting a numerical error less than
$10^{-5}$, illustrating that our numerical
procedure is rather accurate. 

\subsection{Pattern (V): charged ellipse}
\label{sec:charged}

\begin{figure*}[tbh]
	\begin{center}
		\includegraphics[width=2\columnwidth]{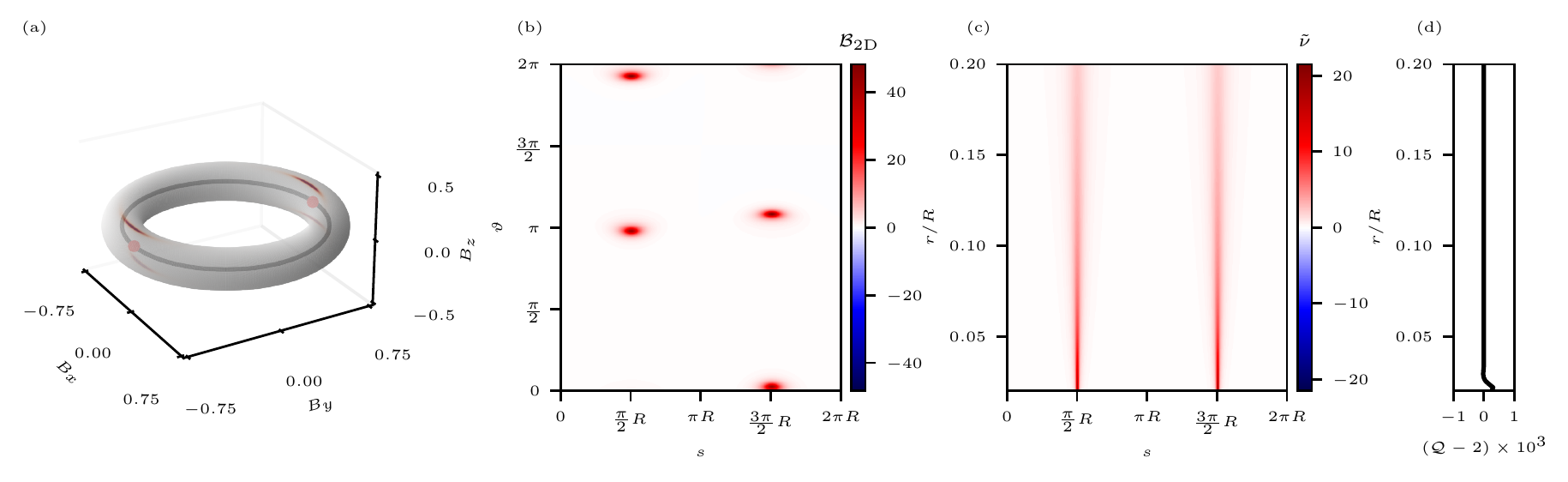}
	\end{center}
	\caption{{\bf Charged degeneracy ellipse
	has two charged points.}
	(a) Berry flux density $\mathcal{B}_n$ (temperature map)
	on the surface of a torus
	surrounding a charged degeneracy circle (black).
	Meridian radius, $r/R=0.2$
    (b) Two-dimensional Berry curvature
    $\mathcal{B}_\text{2D}(s,\vartheta)$
    [Eq.~\eqref{eq:2dberrycurvature}]
    on the pre-image of the torus, i.e., as a function 
    of longitude path length $s$ and meridian angle $\vartheta$. 
    (c) Apparent linear topological charge density $\tilde{\nu}_r(s)$
    [Eq.~\eqref{eq:apparent}],
    as function of longitude path length $s$ and
    meridian radius $r$.
    As $r\to 0$, this function converges to the
    constant zero function (white), showing 
    that the linear topological charge density
    vanishes, $\nu(s) = 0$.
    (d) A benchmark for the numerical integration:
    numerically evaluated ground-state
    Chern number on the torus, 
    as function of the meridian radius $r$.
    Numerical error grows slightly as $r\to 0$,
    but it remains well below $10^{-3}$
    even for the smallest $r$ values considered.
	\label{fig:charged}}
\end{figure*}

Next we consider the charged ellipse
degeneracy pattern (V) in Fig.~\ref{fig:geometries}.
The question is: how is the topological 
charge distributed along the ellipse?
Using the method of the previous subsection, 
we show that the topological charge
is concentrated at two opposite points
of the ellipse, i.e. it is not continuously 
distributed along the ellipse. 

The example parameter set we use consists
of $g$-tensors
\begin{eqnarray} 
\hat{\vec g}_{\text{L,V}}=
\begin{pmatrix}
2&0&0\\
0&2&0\\
0&1&2
\end{pmatrix},
\hspace{6mm}
\hat{\vec g}_{\text{R,V}}=\mathds{1}_{3\times 3},
\end{eqnarray}
and the interaction is described by $J_{\text{V}}=1$ and
\mbox{$\hat{\vec R}_{\text{V}}=\mathds{1}_{3\times 3}$}.
The degeneracy ellipse is a circle 
with radius \mbox{$R=\frac{3}{4}$} again, 
shown as a black line in Fig.~\ref{fig:charged}~a.

Fig.~\ref{fig:charged} shows
(a) the Berry flux density $\mathcal{B}_n$ on a torus
surrounding the degeneracy circle,
(b) the two-dimensional Berry curvature
$\mathcal{B}_\text{2D}(s,\vartheta)$ on
the pre-image $[0,2\pi R)\times [0,2\pi)$ of the torus,
(c) the apparent topological charge density 
$\tilde{\nu}_r(s)$ of the degeneracy circle,
and (d) the numerically evaluated ground-state
Chern number on the torus.

Fig.~\ref{fig:charged}~a and b  
reveal a remarkable difference compared to 
Fig.~\ref{fig:neutral}~a and b:
from Fig.~\ref{fig:charged}~a and b, 
the Berry flux is concentrated in narrow regions
(red spots) in the neighborhoods of two opposite
points of the ellipse. 
Fig.~\ref{fig:charged}~c suggests that
the linear topological charge density,
which corresponds to the plotted data $\tilde{\nu}_r(s)$
in the $r\to 0$ limit,
consists of two Dirac deltas: the degeneracy circle
is neutral in all points except two
discrete points opposite to each other,
each carrying a topological charge of $+1$.
In section~\ref{sec:rank}, this numerical
evidence is supported by analytical 
results.

Fig.~\ref{fig:charged}~d shows that the numerical 
error of the Chern number is below
$10^{-3}$, illustrating the accuracy of our numerical procedure. 
The feature that the error grows as the radius decreases is rather
natural: for smaller radius, the Berry flux gets more focused on a smaller
area, hence our numerical integration using an equidistant grid 
on the pre-image of the torus gets less accurate. 

The direction of the charged points is is $(0,1,0)^{\text{T}}$, as
determined by the Jordan decomposition of $\hat{\vec M}$, according to Eq.~\eqref{eq:jordan5b}. 
From Eq.~\eqref{eq:bnorm}, the position of these points is expressed as: 
\begin{eqnarray}
\vec B_\pm=\pm\frac{3}{4}
\begin{pmatrix}
0\\
1\\
0
\end{pmatrix}.
\end{eqnarray}
This result is in agreement with Fig.~\ref{fig:charged}~c, where the charge density has two 
peaks at \mbox{$s\in\{\frac{\pi}{2}R,\frac{3\pi}{2}R\}$}.

Figure~\ref{fig:charged}~b shows pronounced peaks of the two-dimensional 
Berry curvature. 
These peaks appear because at each charged degeneracy point, there is 
a direction perpendicular to the degeneracy circle in which the
energy splitting grows nonlinearly, and the Berry curvature peaks
in those directions. 
For both degeneracy points, this direction, determined analytically 
using Eq.~\eqref{eq:dyadicgeff2}, is $(0,-1,6)^{\text{T}}$.
This feature appears in the $(s,\vartheta)$ torus of Fig.~\ref{fig:charged}b
as
which appears in Fig.~\ref{fig:charged}~b as the regions where the flux density is high parametrized by the coordinates
\begin{equation}
\begin{split}
(s_1,\vartheta_1)&=\left(\frac{\pi}{2}R,\pi-\tan^{-1}\frac{1}{6}\right),\\
(s_2,\vartheta_2)&=\left(\frac{\pi}{2}R,2\pi-\tan^{-1}\frac{1}{6}\right),\\
(s_3,\vartheta_3)&=\left(\frac{3\pi}{2}R,\pi+\tan^{-1}\frac{1}{6}\right),\\
(s_4,\vartheta_4)&=\left(\frac{3\pi}{2}R,\tan^{-1}\frac{1}{6}\right),
\end{split}
\end{equation}
matching the peaks seen in the numerical data.

To conclude, in this section we provided numerical 
evidence that the neutral degeneracy ellipse, pattern (II) of~\cite{Frank},
has vanishing linear topological charge density,
whereas the charged degeneracy ellipse, pattern (V) of~\cite{Frank}, 
has all its topological charge focused in two opposite points
of the ellipse.
Even though the numerical results are obtained here
for a specific choice of secondary parameters ($g$-tensors, 
exchange strength $J$ and exchange rotation matrix $\hat{\vec R}$), the statements are
general. 
For example, if the secondary parameters are changed with respect
to those in section~\ref{sec:neutral}, such that the resulting matrix $\hat{\vec M}$ still
has the eigenpattern (II), then the degeneracy circle generically deforms into 
an ellipse, but all of its points remain charge-neutral.
Results of section~\ref{sec:charged} are generalized analogously. 
For details, we refer to Appendices~\ref{sec:jordannormalforms}, 
\ref{sec:effectivegtensor}, and~\ref{sec:zerodensity}.

\section{Pattern (IV): continous 
surface charge distribution on an ellipsoid}
\label{sec:ellipsoid}

Consider now the degeneracy pattern (IV) from 
Fig.~\ref{fig:geometries}, the charged ellipsoid. 
Again, we will follow the electrostatics analogy
to determine the surface topological charge distribution 
on this ellipsoid, see also Ref.~\cite{Souza}.
In electrostatics the surface charge density $\sigma(\vec r_S)$
of surface $S$ and the electric field
$\vec E(\vec r)$ created by the surface charge 
density are related by the following formula:
\begin{equation}
    \frac{\sigma(\vec r_S)}{\varepsilon_0}
    = E_n(\vec r_{S+}) - E_n(\vec r_{S-}), 
\end{equation}
where $E_n(\vec r_{S+})$ 
($E_n(\vec r_{S-})$)
is the normal component
of the electric field outside (inside) the surface
at point $\vec r_S$ on the surface. 
Analogously, the surface topological charge density
of the degeneracy ellipsoid $S$ 
is related to the Berry curvature vector
field via
\begin{equation}
\label{eq:surfacecharge}
    \sigma(\vec B_S)
    = \frac{1}{2\pi }
    \left(
        \mathcal{B}_n(\vec B_{S+}) - \mathcal{B}_n(\vec B_{S-})
    \right),
\end{equation}
where $\vec B_S$ is a point of the 
degeneracy surface. 

\begin{figure}[!tbh]
	\begin{center}
		\includegraphics[width=0.9\columnwidth]{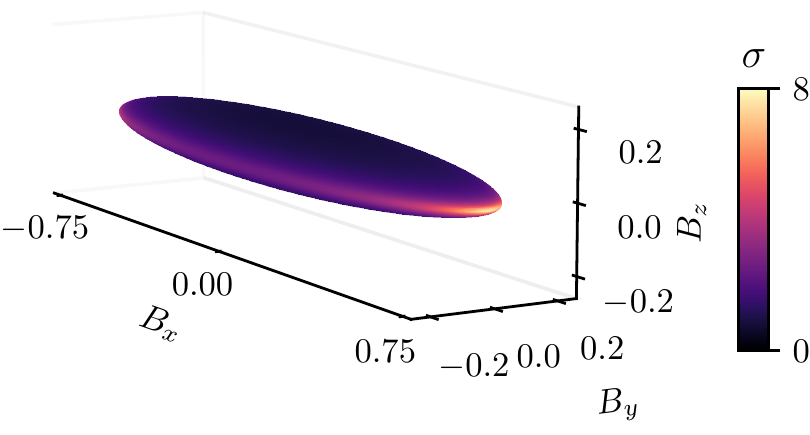}
	\end{center}
	\caption{{\bf Topological surface charge distribution
	on a degeneracy ellipsoid.} 
	See section~\ref{sec:ellipsoid} for 
	parameter values. This topological charge distribution is 
	essentially 
	the same as the electric charge distribution on 
	the surface of a
	charged conducting ellipsoid.
	\label{fig:ellipsoid}}
\end{figure}

Fig.~\ref{fig:ellipsoid} shows this surface topological 
charge distribution $\sigma(\vec B_S)$
for the example parameter set with $g$-tensors
\begin{eqnarray} 
\hat{\vec g}_{\text{L,IV}}=
\begin{pmatrix}
2&0&0\\
0&6&0\\
0&0&18
\end{pmatrix},
\hspace{6mm}
\hat{\vec g}_{\text{R,IV}}=
\begin{pmatrix}
1&0&0\\
0&3&0\\
0&0&9
\end{pmatrix},
\end{eqnarray}
and interaction described by $J_{\text{IV}}=1$ and
$\hat{\vec R}_{\text{IV}}=\mathds{1}_{3\times 3}$.
In contrast to the result of section~\ref{sec:charged},
here we observe a continuous charge distribution. 
Figure~\ref{fig:ellipsoid} is obtained from
our general result for the surface topological charge distribution, 
which reads
\begin{equation}
\label{eq:surfacechargedensity}
    \sigma(\vec B_S)=\frac{a\det\hat{\vec g}_\text{R}}{\pi(a+1)|\hat{\vec g}_\text{R}\hat{\vec g}_\text{R}^{\text T}\vec B_S|}.
\end{equation}
Here, $a$ is the diagonal element of the Jordan normal 
form of the matrix $\hat{\vec M}$, which
is $a\cdot\mathds{1}_{3\times 3}$~\cite{Frank}.
Equation \eqref{eq:surfacechargedensity}
is derived in Appendix~\ref{sec:surfacederivation}.

Interestingly, the surface charge density in 
Eq.~\eqref{eq:surfacechargedensity} has
the same functional form as the 
electrical charge distribution of an
electrically charged conducting ellipsoid~\cite{Curtright2020charge}.
Figure~\ref{fig:ellipsoid} exhibits the 
curvature effect known from electrostatics: 
the greater the curvature of the surface,
the greater 
the topological charge density.
A further similarity is that 
the Berry curvature inside the ellipsoid
is zero, similarly to the electric field inside
a charged ideal conductor. 
A difference, however, is that
the Berry curvature in our example exits the surface
radially with respect to the origin (i.e., it is proportional to $\vec B/B$), 
in contrast to the electric field 
which exits the conductor's surface in the surface normal direction
(i.e., it is proportional to $\vec n(\vec B)$). 
Another difference is that the curl of the Berry curvature
is nonzero, however, the curl of the electric field induced
by the charged conductor vanishes.

Given a degeneracy surface in a three-dimensional
parameter space, is it a generic feature that 
it carries a continuous topological surface 
charge density?
Here we argue that it is. 
Such a degeneracy surface divides the parameter space
to two disjoint regions - in our example, the inside
and the outside of the ellipsoid. 
The ground state changes continuously in both regions 
as the function of the parameters, generically implying
a nonzero and continuous Berry curvature 
vector fields
in both regions separately.
But at the degeneracy surface, the ground state
changes suddenly - in our example from 
a singlet-like state at the inside and a triplet-like
state at the outside - and hence the Berry curvature
also jumps, leading to a finite surface charge
density according to Eq.~\eqref{eq:surfacecharge}.

\section{Discussion}
\label{sec:discussion}

\subsection{Topological charge density 
vanishes for rank-2 points of degeneracy lines}
\label{sec:rank}

In this section, we outline analytical 
results that support the numerical 
evidence of topological charge distributions
studied in section~\ref{sec:linear}.
To this end, we define the \emph{rank} of 
degeneracy points, establish the ranks of the
degeneracy points forming the linear degeneracy
patterns studied in section~\ref{sec:linear}, 
and relate the rank of a degeneracy point
to the topological charge of that point. 
In particular, we find that a rank-2 degeneracy
point embedded in a linear degeneracy pattern 
carries no topological charge. 

The \emph{effective $g$-tensor} $\hat{\vec g}_{\text{eff}}(\vec B_0)$ of 
a degeneracy point $\vec B_0$
is a $3\times 3$ real matrix that
characterizes the Hamiltonian
in the parameter-space vicinity of the 
degeneracy point $\vec B_0$, focusing on the two
levels that are degenerate at the degeneracy point. 
Formally, we introduce the \emph{relative parameter vector}
$\delta \vec B = \vec B- \vec B_0$
measured from the degeneracy point, 
project the Hamiltonian $H(\vec B_0 + \delta \vec B)$
to the two-dimensional ground-state subspace 
of the degeneracy point $\vec B_0$
using an arbitrary orthonormal basis $(\ket{0},\ket{1})$, 
and express that projected Hamiltonian
in terms of Pauli matrices $\vec \tau = (\tau_x,\tau_y,\tau_z)$, e.g., $\tau_z =\frac{1}{2}\left( \ket{0}\bra{0}-\ket{1}\bra{1}\right)$, 
leading to the form~\cite{Frank}
\begin{equation}\label{eq:Heff}
    H_\text{p}(\delta \vec B) 
    = \delta \vec B \cdot \hat{\vec g}_{\text{eff}}(\vec B_0) \,  \vec \tau.
\end{equation}
Here, we have omitted the projected $H_\text{p}(\vec B_0)$, since
it is proportional to the $2\times 2$ unit matrix, owing to the degeneracy of the 
relevant two-dimensional subspace at $\vec B_0$. 
By the \emph{rank} of a degeneracy point $\vec B_0$, we mean 
the matrix rank of the effective $g$-tensor of that
degeneracy point.
Even though the effective $g$-tensor depends on the choice
of the basis $(\ket{0},\ket{1})$, its 
determinant and rank do not.

In Appendix~\ref{sec:neutraleffectiveg},
we show that all degeneracy points of the neutral degeneracy 
ellipse are rank 2.
Furthermore, in Appendix~\ref{sec:chargedeffectiveg}
we show that in the charged degeneracy
ellipse, the two points where the Berry flux 
density is concentrated in Fig.~\ref{fig:charged}
are rank 1, and all other degeneracy points are rank 2.
Finally, in Appendix~\ref{sec:zerodensity}, we prove
that the linear topological charge density at
a rank-2 degeneracy point of a line degeneracy
is zero, supporting the numerical 
evidence seen in Figs.~\ref{fig:neutral}c and
Figs.~\ref{fig:charged}c.
We also observe that the electrostatic analogy is not perfect:
while the local charge density vanishes, hence there is no source of the Berry curvature on rank-2 degeneracy lines, there is always a linelike $\pi$ flux tube along the degeneracy, for details see Appendix~\ref{sec:zerodensity}.

It is tempting to think about the charged 
ellipse degeneracy pattern (V) as a result of 
fine-tuning of pattern (II): upon
tuning the secondary parameters ($g$-tensors, exchange
parameters), two charged points of pattern (II) 
merge with the neutral degeneracy ellipse of pattern (II), 
forming the charged degeneracy ellipse of pattern (V). 
(Note the related discussion on the conversion 
between Weyl point and nodal lines in 
band structures~\cite{SunPRL2018}.)
This picture is reinforced by the fact that
pattern (V) is less stable than pattern (II), 
signalled by their stability codimensions 
4 and 3, respectively (see Table I. of Ref.~\onlinecite{Frank}). 
Therefore, we conjecture that it is the generic behavior
of two-fold degeneracy lines that their points have
vanishing linear topological charge density. 
In other words, we conjecture that for $N=3$, 
if a twofold line degeneracy has a point or segment with
a nonzero linear topological charge density, then 
an infinitesimal perturbation, 
which preserves the line degeneracy but is generic
otherwise, will separate
the charge from the line degeneracy and render the latter 
locally neutral. 

\subsection{Symmetries can stabilize non-generic degeneracy 
patterns}

Non-generic band degeneracy points and
patterns in solids can be stabilized
by the presence of symmetries~\cite{ChenFang_multiweyl,ChenFangChinese2016}.
Here, we show that interacting spin systems are similar:
non-generic degeneracy points can be stabilized
by symmetries.

We focus on a special case, when the two-spin 
system described by Eq.~\eqref{eq:hamiltonian}
has $C_{3v}$ symmetry, and leave it for future
work to explore further symmetry groups.
We show that in this case, pattern (II) with two Weyl points
and the neutral degeneracy circle is stabilized, even though
it is unstable (codimension 3, see Table I in Ref.~\onlinecite{Frank}) 
without the symmetry constraint. 

Consider the case when the Hamiltonian is invariant under the isometries
of the group $C_{3v}$, 
which are generated by the threefold rotation $\mathcal{R}$ around
the $z$ axis, and the reflection on the $xz$ plane, $\mathcal{M}$.
These isometries are represented by the $3\times 3$ matrices
\begin{equation}
    \mathcal{R} = 
    \left(
        \begin{array}{ccc}
            \cos\frac{2\pi}{3} & -\sin\frac{2\pi}{3} & 0 \\
            \sin\frac{2\pi}{3} & \cos\frac{2\pi}{3} & 0 \\
            0 & 0 & 1
        \end{array}
    \right),
\end{equation}
and
\begin{equation}
    \mathcal{M} = 
        \left(
        \begin{array}{ccc}
            -1 & 0 & 0 \\
            0  & 1 & 0 \\
            0 & 0 & -1
        \end{array}
    \right),
\end{equation}
on the pseudovectors (or axial vectors) appearing
in the Hamiltonian of Eq.~\eqref{eq:hamiltonian}, 
i.e., the magnetic field
$\vec B$ and the electron spins $\vec S_L$ and $\vec S_R$.

The presence of spatial symmetries is 
formalized
as 
$H(\vec B,\vec S_L, \vec S_R) = 
H(\mathcal{R}\vec B,\mathcal{R}\vec S_L, \mathcal{R}\vec S_R)$
and
$H(\vec B,\vec S_L, \vec S_R) = 
H(\mathcal{M}\vec B,\mathcal{M}\vec S_L, \mathcal{M}\vec S_R)$.
These conditions restrict
the forms of $\hat{\vec g}_L$, $\hat{\vec g}_R$, and $\vec{R}$,
in particular, 
$\hat{\vec g}_L = \text{diag}(g_{Lx},g_{Lx},g_{Lz})$,
$\hat{\vec g}_R = \text{diag}(g_{Rx},g_{Rx},g_{Rz})$,
and 
$\vec{R} = \mathds{1}_{3\times 3}$ or
$\vec{R} = \text{diag}(-1,-1,1)$.
For concreteness, we still assume 
positive determinants for 
the $g$-tensors, which implies $g_{Lz},g_{Rz} > 0$.

Combining these symmetry constraints 
with the definition in Eq.~\eqref{eq:matrixm} leads to
$\hat{\vec M} = \text{diag}(a,a,b)$,
where $b>0$ and $a$ might be either positive or negative.
This matrix $\hat{\vec M}$ is its own Jordan normal form, 
so we can directly apply the Jordan classification in 
Table I of Ref.~\onlinecite{Frank} to determine the degeneracy patterns
arising in the magnetic parameter space. 
For $a<0$, we find eigenpattern (VII),
which has two  Weyl points. 
For $a>0$, we find eigenpattern (II), with two
Weyl points and a neutral ellipse (circle, in this case), 
as studied in section~\ref{sec:neutral}.
Remarkably, as long as the $C_{3v}$ symmetry is intact, 
the neutral ellipse survives without fine-tuning. 
In other words, the degeneracy pattern (II), 
which is unstable in the absence of symmetries, 
and hence is characterized by a positive codimension, becomes stable 
with zero codimension in the presence of $C_{3v}$ symmetry.

\section{Conclusions}
\label{sec:conclusions}

We have exemplified the concepts of linear topological charge 
density and surface topological charge density
through the example of a simple parameter-dependent quantum system, 
the spin-orbit-coupled two-spin problem where the
paramters are the Cartesian components of the magnetic field
acting on the spins. 
We have shown that the neutral degeneracy ellipse has
vanishing topological charge density in all of its points,
whereas the charged degeneracy ellipse has a charge 
distribution that is concentrated in two opposite points in 
the magnetic-field parameter space. 
Moreover, we have shown that the surface topological charge density
of the degeneracy ellipse is continuous, and this 
charge density is identical to the surface charge density of 
a charged conducting ellipsoid. 
We have also shown that if the two-spin system has
certain spatial symmetries, then this can stabilize
an otherwise unstable, non-generic degeneracy pattern, 
e.g., a neutral circle. 

The topological features described in this work have
numerous physical consequences, e.g., they determine
the experimentally measurable Berry 
curvature~\cite{Schroer_chern,Roushan}, 
and also determine dynamical properties, such as 
paramagnetic resonance~\cite{Koppens-esr} 
or Landau--Zener-type 
processes~\cite{Petta_beamsplitter,Tanttu}.
Hence, we expect that our findings are testable in
few-spin experiments, e.g., using quantum
dots~\cite{Scherubl,VeldhorstPRB,Frank}, 
molecular magnets~\cite{Garg,Wernsdorfer,Bruno}
or adatoms on metallic surfaces~\cite{Wiesendanger,Spinelli}.

\section*{Author contributions}
Gy.~F. and A.~P. formulated the project
and wrote the initial draft of the manuscript. 
A.~P. acquired funding, and managed the project.
Gy.~F. performed analytical and numerical calculations with assistance from A.~P. and D.~V..
G.~P. and P.~V. consulted on differential-geometric aspects of the work.
Gy.~F. and D.~V. produced the figures.
All authors discussed the results and took part in writing the manuscript.

\acknowledgments
We acknowledge fruitful discussions with G. Zaránd. 
This work was supported by the National Research 
Development and Innovation Office of Hungary within the 
Quantum Technology National Excellence Program (Project 
No.~2017-1.2.1-NKP-2017-00001), under OTKA Grants 124723,
132146, and under the BME Nanotechnology and Materials 
Science TKP2020 IE grant  (BME IE-NAT TKP2020).
D.~V. was supported by NWO VIDI grant 680-47-53, the Swedish Research Council (VR) and the Knut and Alice Wallenberg Foundation.

\bibliography{Gy_Frank}

\begin{thebibliography}{44}%
\makeatletter
\providecommand \@ifxundefined [1]{%
 \@ifx{#1\undefined}
}%
\providecommand \@ifnum [1]{%
 \ifnum #1\expandafter \@firstoftwo
 \else \expandafter \@secondoftwo
 \fi
}%
\providecommand \@ifx [1]{%
 \ifx #1\expandafter \@firstoftwo
 \else \expandafter \@secondoftwo
 \fi
}%
\providecommand \natexlab [1]{#1}%
\providecommand \enquote  [1]{``#1''}%
\providecommand \bibnamefont  [1]{#1}%
\providecommand \bibfnamefont [1]{#1}%
\providecommand \citenamefont [1]{#1}%
\providecommand \href@noop [0]{\@secondoftwo}%
\providecommand \href [0]{\begingroup \@sanitize@url \@href}%
\providecommand \@href[1]{\@@startlink{#1}\@@href}%
\providecommand \@@href[1]{\endgroup#1\@@endlink}%
\providecommand \@sanitize@url [0]{\catcode `\\12\catcode `\$12\catcode
  `\&12\catcode `\#12\catcode `\^12\catcode `\_12\catcode `\%12\relax}%
\providecommand \@@startlink[1]{}%
\providecommand \@@endlink[0]{}%
\providecommand \url  [0]{\begingroup\@sanitize@url \@url }%
\providecommand \@url [1]{\endgroup\@href {#1}{\urlprefix }}%
\providecommand \urlprefix  [0]{URL }%
\providecommand \Eprint [0]{\href }%
\providecommand \doibase [0]{https://doi.org/}%
\providecommand \selectlanguage [0]{\@gobble}%
\providecommand \bibinfo  [0]{\@secondoftwo}%
\providecommand \bibfield  [0]{\@secondoftwo}%
\providecommand \translation [1]{[#1]}%
\providecommand \BibitemOpen [0]{}%
\providecommand \bibitemStop [0]{}%
\providecommand \bibitemNoStop [0]{.\EOS\space}%
\providecommand \EOS [0]{\spacefactor3000\relax}%
\providecommand \BibitemShut  [1]{\csname bibitem#1\endcsname}%
\let\auto@bib@innerbib\@empty
\bibitem [{\citenamefont {Herring}(1937)}]{Herring}%
  \BibitemOpen
  \bibfield  {author} {\bibinfo {author} {\bibfnamefont {C.}~\bibnamefont
  {Herring}},\ }\bibfield  {title} {\bibinfo {title} {Accidental degeneracy in
  the energy bands of crystals},\ }\href
  {https://doi.org/10.1103/PhysRev.52.365} {\bibfield  {journal} {\bibinfo
  {journal} {Phys. Rev.}\ }\textbf {\bibinfo {volume} {52}},\ \bibinfo {pages}
  {365} (\bibinfo {year} {1937})}\BibitemShut {NoStop}%
\bibitem [{\citenamefont {Berry}(1984)}]{Berry}%
  \BibitemOpen
  \bibfield  {author} {\bibinfo {author} {\bibfnamefont {M.~V.}\ \bibnamefont
  {Berry}},\ }\bibfield  {title} {\bibinfo {title} {Quantal phase factors
  accompanying adiabatic changes},\ }\href
  {https://doi.org/10.1098/rspa.1984.0023} {\bibfield  {journal} {\bibinfo
  {journal} {Proceedings of the Royal Society of London A: Mathematical,
  Physical and Engineering Sciences}\ }\textbf {\bibinfo {volume} {392}},\
  \bibinfo {pages} {45} (\bibinfo {year} {1984})}\BibitemShut {NoStop}%
\bibitem [{\citenamefont {Hasan}\ and\ \citenamefont
  {Kane}(2010)}]{HasanRMP2010}%
  \BibitemOpen
  \bibfield  {author} {\bibinfo {author} {\bibfnamefont {M.~Z.}\ \bibnamefont
  {Hasan}}\ and\ \bibinfo {author} {\bibfnamefont {C.~L.}\ \bibnamefont
  {Kane}},\ }\bibfield  {title} {\bibinfo {title} {Colloquium: Topological
  insulators},\ }\href {https://doi.org/10.1103/RevModPhys.82.3045} {\bibfield
  {journal} {\bibinfo  {journal} {Rev. Mod. Phys.}\ }\textbf {\bibinfo {volume}
  {82}},\ \bibinfo {pages} {3045} (\bibinfo {year} {2010})}\BibitemShut
  {NoStop}%
\bibitem [{\citenamefont {Armitage}\ \emph {et~al.}(2018)\citenamefont
  {Armitage}, \citenamefont {Mele},\ and\ \citenamefont
  {Vishwanath}}]{Armitage}%
  \BibitemOpen
  \bibfield  {author} {\bibinfo {author} {\bibfnamefont {N.~P.}\ \bibnamefont
  {Armitage}}, \bibinfo {author} {\bibfnamefont {E.~J.}\ \bibnamefont {Mele}},\
  and\ \bibinfo {author} {\bibfnamefont {A.}~\bibnamefont {Vishwanath}},\
  }\bibfield  {title} {\bibinfo {title} {{Weyl and Dirac semimetals in
  three-dimensional solids}},\ }\href
  {https://doi.org/10.1103/RevModPhys.90.015001} {\bibfield  {journal}
  {\bibinfo  {journal} {Rev. Mod. Phys.}\ }\textbf {\bibinfo {volume} {90}},\
  \bibinfo {pages} {015001} (\bibinfo {year} {2018})}\BibitemShut {NoStop}%
\bibitem [{\citenamefont {Asb\'oth}\ \emph {et~al.}(2016)\citenamefont
  {Asb\'oth}, \citenamefont {Oroszl\'any},\ and\ \citenamefont
  {P\'alyi}}]{Asboth}%
  \BibitemOpen
  \bibfield  {author} {\bibinfo {author} {\bibfnamefont {J.~K.}\ \bibnamefont
  {Asb\'oth}}, \bibinfo {author} {\bibfnamefont {L.}~\bibnamefont
  {Oroszl\'any}},\ and\ \bibinfo {author} {\bibfnamefont {A.}~\bibnamefont
  {P\'alyi}},\ }\href@noop {} {\emph {\bibinfo {title} {A Short Course on
  Topological Insulators}}}\ (\bibinfo  {publisher} {Springer},\ \bibinfo
  {address} {Heidelberg},\ \bibinfo {year} {2016})\BibitemShut {NoStop}%
\bibitem [{\citenamefont {Riwar}\ \emph {et~al.}(2016)\citenamefont {Riwar},
  \citenamefont {Houzet}, \citenamefont {Meyer},\ and\ \citenamefont
  {Nazarov}}]{Riwar}%
  \BibitemOpen
  \bibfield  {author} {\bibinfo {author} {\bibfnamefont {R.-P.}\ \bibnamefont
  {Riwar}}, \bibinfo {author} {\bibfnamefont {M.}~\bibnamefont {Houzet}},
  \bibinfo {author} {\bibfnamefont {J.~S.}\ \bibnamefont {Meyer}},\ and\
  \bibinfo {author} {\bibfnamefont {Y.~V.}\ \bibnamefont {Nazarov}},\
  }\bibfield  {title} {\bibinfo {title} {{Multi-terminal Josephson junctions as
  topological matter}},\ }\href@noop {} {\bibfield  {journal} {\bibinfo
  {journal} {Nature Communications}\ }\textbf {\bibinfo {volume} {7}},\
  \bibinfo {pages} {11167} (\bibinfo {year} {2016})}\BibitemShut {NoStop}%
\bibitem [{\citenamefont {Scher{\"u}bl}\ \emph {et~al.}(2019)\citenamefont
  {Scher{\"u}bl}, \citenamefont {P{\'a}lyi}, \citenamefont {Frank},
  \citenamefont {Luk{\'a}cs}, \citenamefont {F{\"u}l{\"o}p}, \citenamefont
  {F{\"u}l{\"o}p}, \citenamefont {Nyg{\aa}rd}, \citenamefont {Watanabe},
  \citenamefont {Taniguchi}, \citenamefont {Zar{\'a}nd},\ and\ \citenamefont
  {Csonka}}]{Scherubl}%
  \BibitemOpen
  \bibfield  {author} {\bibinfo {author} {\bibfnamefont {Z.}~\bibnamefont
  {Scher{\"u}bl}}, \bibinfo {author} {\bibfnamefont {A.}~\bibnamefont
  {P{\'a}lyi}}, \bibinfo {author} {\bibfnamefont {G.}~\bibnamefont {Frank}},
  \bibinfo {author} {\bibfnamefont {I.~E.}\ \bibnamefont {Luk{\'a}cs}},
  \bibinfo {author} {\bibfnamefont {G.}~\bibnamefont {F{\"u}l{\"o}p}}, \bibinfo
  {author} {\bibfnamefont {B.}~\bibnamefont {F{\"u}l{\"o}p}}, \bibinfo {author}
  {\bibfnamefont {J.}~\bibnamefont {Nyg{\aa}rd}}, \bibinfo {author}
  {\bibfnamefont {K.}~\bibnamefont {Watanabe}}, \bibinfo {author}
  {\bibfnamefont {T.}~\bibnamefont {Taniguchi}}, \bibinfo {author}
  {\bibfnamefont {G.}~\bibnamefont {Zar{\'a}nd}},\ and\ \bibinfo {author}
  {\bibfnamefont {S.}~\bibnamefont {Csonka}},\ }\bibfield  {title} {\bibinfo
  {title} {Observation of spin--orbit coupling induced {Weyl} points in a
  two-electron double quantum dot},\ }\href@noop {} {\bibfield  {journal}
  {\bibinfo  {journal} {Communications Physics}\ }\textbf {\bibinfo {volume}
  {2}},\ \bibinfo {pages} {108} (\bibinfo {year} {2019})}\BibitemShut {NoStop}%
\bibitem [{\citenamefont {Wernsdorfer}\ and\ \citenamefont
  {Sessoli}(1999)}]{Wernsdorfer}%
  \BibitemOpen
  \bibfield  {author} {\bibinfo {author} {\bibfnamefont {W.}~\bibnamefont
  {Wernsdorfer}}\ and\ \bibinfo {author} {\bibfnamefont {R.}~\bibnamefont
  {Sessoli}},\ }\bibfield  {title} {\bibinfo {title} {Quantum phase
  interference and parity effects in magnetic molecular clusters},\ }\href
  {https://doi.org/10.1126/science.284.5411.133} {\bibfield  {journal}
  {\bibinfo  {journal} {Science}\ }\textbf {\bibinfo {volume} {284}},\ \bibinfo
  {pages} {133} (\bibinfo {year} {1999})}\BibitemShut {NoStop}%
\bibitem [{\citenamefont {Bruno}(2006)}]{Bruno}%
  \BibitemOpen
  \bibfield  {author} {\bibinfo {author} {\bibfnamefont {P.}~\bibnamefont
  {Bruno}},\ }\bibfield  {title} {\bibinfo {title} {{Berry} phase, topology,
  and degeneracies in quantum nanomagnets},\ }\href
  {https://doi.org/10.1103/PhysRevLett.96.117208} {\bibfield  {journal}
  {\bibinfo  {journal} {Phys. Rev. Lett.}\ }\textbf {\bibinfo {volume} {96}},\
  \bibinfo {pages} {117208} (\bibinfo {year} {2006})}\BibitemShut {NoStop}%
\bibitem [{\citenamefont {Gritsev}\ and\ \citenamefont
  {Polkovnikov}(2012)}]{Gritsev}%
  \BibitemOpen
  \bibfield  {author} {\bibinfo {author} {\bibfnamefont {V.}~\bibnamefont
  {Gritsev}}\ and\ \bibinfo {author} {\bibfnamefont {A.}~\bibnamefont
  {Polkovnikov}},\ }\bibfield  {title} {\bibinfo {title} {Dynamical quantum
  {Hall} effect in the parameter space},\ }\href
  {https://doi.org/10.1073/pnas.1116693109} {\bibfield  {journal} {\bibinfo
  {journal} {Proceedings of the National Academy of Sciences}\ }\textbf
  {\bibinfo {volume} {109}},\ \bibinfo {pages} {6457} (\bibinfo {year}
  {2012})}\BibitemShut {NoStop}%
\bibitem [{\citenamefont {Frank}\ \emph {et~al.}(2020)\citenamefont {Frank},
  \citenamefont {Scher{\"u}bl}, \citenamefont {Csonka}, \citenamefont
  {Zar{\'a}nd},\ and\ \citenamefont {P{\'a}lyi}}]{Frank}%
  \BibitemOpen
  \bibfield  {author} {\bibinfo {author} {\bibfnamefont {G.}~\bibnamefont
  {Frank}}, \bibinfo {author} {\bibfnamefont {Z.}~\bibnamefont {Scher{\"u}bl}},
  \bibinfo {author} {\bibfnamefont {S.}~\bibnamefont {Csonka}}, \bibinfo
  {author} {\bibfnamefont {G.}~\bibnamefont {Zar{\'a}nd}},\ and\ \bibinfo
  {author} {\bibfnamefont {A.}~\bibnamefont {P{\'a}lyi}},\ }\bibfield  {title}
  {\bibinfo {title} {Magnetic degeneracy points in interacting two-spin
  systems: Geometrical patterns, topological charge distributions, and their
  stability},\ }\href@noop {} {\bibfield  {journal} {\bibinfo  {journal}
  {Physical Review B}\ }\textbf {\bibinfo {volume} {101}},\ \bibinfo {pages}
  {245409} (\bibinfo {year} {2020})}\BibitemShut {NoStop}%
\bibitem [{\citenamefont {von Neumann}\ and\ \citenamefont
  {Wigner}(1929)}]{Neumann}%
  \BibitemOpen
  \bibfield  {author} {\bibinfo {author} {\bibfnamefont {J.}~\bibnamefont {von
  Neumann}}\ and\ \bibinfo {author} {\bibfnamefont {E.~P.}\ \bibnamefont
  {Wigner}},\ }\bibfield  {title} {\bibinfo {title} {{\"Uber das Verhalten von
  Eigenwerten bei adiabatischen Prozessen}},\ }\href@noop {} {\bibfield
  {journal} {\bibinfo  {journal} {Physikalische Zeitschrift}\ }\textbf
  {\bibinfo {volume} {30}},\ \bibinfo {pages} {467} (\bibinfo {year}
  {1929})}\BibitemShut {NoStop}%
\bibitem [{\citenamefont {Arnold}(1995)}]{ArnoldSelMath1995}%
  \BibitemOpen
  \bibfield  {author} {\bibinfo {author} {\bibfnamefont {V.~I.}\ \bibnamefont
  {Arnold}},\ }\bibfield  {title} {\bibinfo {title} {Remarks on eigenvalues and
  eigenvectors of {Hermitian} matrices, {Berry} phase, adiabatic connections
  and quantum {Hall} effect},\ }\href@noop {} {\bibfield  {journal} {\bibinfo
  {journal} {Selecta Mathematica}\ }\textbf {\bibinfo {volume} {1}} (\bibinfo
  {year} {1995})}\BibitemShut {NoStop}%
\bibitem [{\citenamefont {Simon}(1983)}]{SimonPRL1983}%
  \BibitemOpen
  \bibfield  {author} {\bibinfo {author} {\bibfnamefont {B.}~\bibnamefont
  {Simon}},\ }\bibfield  {title} {\bibinfo {title} {Holonomy, the quantum
  adiabatic theorem, and {Berry}'s phase},\ }\href
  {https://doi.org/10.1103/PhysRevLett.51.2167} {\bibfield  {journal} {\bibinfo
   {journal} {Phys. Rev. Lett.}\ }\textbf {\bibinfo {volume} {51}},\ \bibinfo
  {pages} {2167} (\bibinfo {year} {1983})}\BibitemShut {NoStop}%
\bibitem [{\citenamefont {Fang}\ \emph {et~al.}(2012)\citenamefont {Fang},
  \citenamefont {Gilbert}, \citenamefont {Dai},\ and\ \citenamefont
  {Bernevig}}]{ChenFang_multiweyl}%
  \BibitemOpen
  \bibfield  {author} {\bibinfo {author} {\bibfnamefont {C.}~\bibnamefont
  {Fang}}, \bibinfo {author} {\bibfnamefont {M.~J.}\ \bibnamefont {Gilbert}},
  \bibinfo {author} {\bibfnamefont {X.}~\bibnamefont {Dai}},\ and\ \bibinfo
  {author} {\bibfnamefont {B.~A.}\ \bibnamefont {Bernevig}},\ }\bibfield
  {title} {\bibinfo {title} {Multi-{Weyl} topological semimetals stabilized by
  point group symmetry},\ }\href
  {https://doi.org/10.1103/PhysRevLett.108.266802} {\bibfield  {journal}
  {\bibinfo  {journal} {Phys. Rev. Lett.}\ }\textbf {\bibinfo {volume} {108}},\
  \bibinfo {pages} {266802} (\bibinfo {year} {2012})}\BibitemShut {NoStop}%
\bibitem [{\citenamefont {Yan}\ and\ \citenamefont {Wang}(2017)}]{ZhongboYan}%
  \BibitemOpen
  \bibfield  {author} {\bibinfo {author} {\bibfnamefont {Z.}~\bibnamefont
  {Yan}}\ and\ \bibinfo {author} {\bibfnamefont {Z.}~\bibnamefont {Wang}},\
  }\bibfield  {title} {\bibinfo {title} {Floquet multi-{Weyl} points in
  crossing-nodal-line semimetals},\ }\href
  {https://doi.org/10.1103/PhysRevB.96.041206} {\bibfield  {journal} {\bibinfo
  {journal} {Phys. Rev. B}\ }\textbf {\bibinfo {volume} {96}},\ \bibinfo
  {pages} {041206} (\bibinfo {year} {2017})}\BibitemShut {NoStop}%
\bibitem [{\citenamefont {Ahn}\ \emph {et~al.}(2017)\citenamefont {Ahn},
  \citenamefont {Mele},\ and\ \citenamefont {Min}}]{Ahn}%
  \BibitemOpen
  \bibfield  {author} {\bibinfo {author} {\bibfnamefont {S.}~\bibnamefont
  {Ahn}}, \bibinfo {author} {\bibfnamefont {E.~J.}\ \bibnamefont {Mele}},\ and\
  \bibinfo {author} {\bibfnamefont {H.}~\bibnamefont {Min}},\ }\bibfield
  {title} {\bibinfo {title} {Optical conductivity of multi-{Weyl} semimetals},\
  }\href {https://doi.org/10.1103/PhysRevB.95.161112} {\bibfield  {journal}
  {\bibinfo  {journal} {Phys. Rev. B}\ }\textbf {\bibinfo {volume} {95}},\
  \bibinfo {pages} {161112} (\bibinfo {year} {2017})}\BibitemShut {NoStop}%
\bibitem [{\citenamefont {Huang}\ \emph {et~al.}(2017)\citenamefont {Huang},
  \citenamefont {Zhou},\ and\ \citenamefont {Shen}}]{ZeMinHuang}%
  \BibitemOpen
  \bibfield  {author} {\bibinfo {author} {\bibfnamefont {Z.-M.}\ \bibnamefont
  {Huang}}, \bibinfo {author} {\bibfnamefont {J.}~\bibnamefont {Zhou}},\ and\
  \bibinfo {author} {\bibfnamefont {S.-Q.}\ \bibnamefont {Shen}},\ }\bibfield
  {title} {\bibinfo {title} {Topological responses from chiral anomaly in
  multi-{Weyl} semimetals},\ }\href
  {https://doi.org/10.1103/PhysRevB.96.085201} {\bibfield  {journal} {\bibinfo
  {journal} {Phys. Rev. B}\ }\textbf {\bibinfo {volume} {96}},\ \bibinfo
  {pages} {085201} (\bibinfo {year} {2017})}\BibitemShut {NoStop}%
\bibitem [{\citenamefont {B\'eri}(2010)}]{BeriPRB2010}%
  \BibitemOpen
  \bibfield  {author} {\bibinfo {author} {\bibfnamefont {B.}~\bibnamefont
  {B\'eri}},\ }\bibfield  {title} {\bibinfo {title} {Topologically stable
  gapless phases of time-reversal-invariant superconductors},\ }\href
  {https://doi.org/10.1103/PhysRevB.81.134515} {\bibfield  {journal} {\bibinfo
  {journal} {Phys. Rev. B}\ }\textbf {\bibinfo {volume} {81}},\ \bibinfo
  {pages} {134515} (\bibinfo {year} {2010})}\BibitemShut {NoStop}%
\bibitem [{\citenamefont {Carter}\ \emph {et~al.}(2012)\citenamefont {Carter},
  \citenamefont {Shankar}, \citenamefont {Zeb},\ and\ \citenamefont
  {Kee}}]{CarterPRB2012}%
  \BibitemOpen
  \bibfield  {author} {\bibinfo {author} {\bibfnamefont {J.-M.}\ \bibnamefont
  {Carter}}, \bibinfo {author} {\bibfnamefont {V.~V.}\ \bibnamefont {Shankar}},
  \bibinfo {author} {\bibfnamefont {M.~A.}\ \bibnamefont {Zeb}},\ and\ \bibinfo
  {author} {\bibfnamefont {H.-Y.}\ \bibnamefont {Kee}},\ }\bibfield  {title}
  {\bibinfo {title} {Semimetal and topological insulator in perovskite
  iridates},\ }\href {https://doi.org/10.1103/PhysRevB.85.115105} {\bibfield
  {journal} {\bibinfo  {journal} {Phys. Rev. B}\ }\textbf {\bibinfo {volume}
  {85}},\ \bibinfo {pages} {115105} (\bibinfo {year} {2012})}\BibitemShut
  {NoStop}%
\bibitem [{\citenamefont {Fang}\ \emph {et~al.}(2015)\citenamefont {Fang},
  \citenamefont {Chen}, \citenamefont {Kee},\ and\ \citenamefont
  {Fu}}]{ChenFang}%
  \BibitemOpen
  \bibfield  {author} {\bibinfo {author} {\bibfnamefont {C.}~\bibnamefont
  {Fang}}, \bibinfo {author} {\bibfnamefont {Y.}~\bibnamefont {Chen}}, \bibinfo
  {author} {\bibfnamefont {H.-Y.}\ \bibnamefont {Kee}},\ and\ \bibinfo {author}
  {\bibfnamefont {L.}~\bibnamefont {Fu}},\ }\bibfield  {title} {\bibinfo
  {title} {Topological nodal line semimetals with and without spin-orbital
  coupling},\ }\href {https://doi.org/10.1103/PhysRevB.92.081201} {\bibfield
  {journal} {\bibinfo  {journal} {Phys. Rev. B}\ }\textbf {\bibinfo {volume}
  {92}},\ \bibinfo {pages} {081201} (\bibinfo {year} {2015})}\BibitemShut
  {NoStop}%
\bibitem [{\citenamefont {Wu}\ \emph {et~al.}(2018)\citenamefont {Wu},
  \citenamefont {Liu}, \citenamefont {Li}, \citenamefont {Zhong}, \citenamefont
  {Yu}, \citenamefont {Sheng}, \citenamefont {Zhao},\ and\ \citenamefont
  {Yang}}]{WeikangWu}%
  \BibitemOpen
  \bibfield  {author} {\bibinfo {author} {\bibfnamefont {W.}~\bibnamefont
  {Wu}}, \bibinfo {author} {\bibfnamefont {Y.}~\bibnamefont {Liu}}, \bibinfo
  {author} {\bibfnamefont {S.}~\bibnamefont {Li}}, \bibinfo {author}
  {\bibfnamefont {C.}~\bibnamefont {Zhong}}, \bibinfo {author} {\bibfnamefont
  {Z.-M.}\ \bibnamefont {Yu}}, \bibinfo {author} {\bibfnamefont {X.-L.}\
  \bibnamefont {Sheng}}, \bibinfo {author} {\bibfnamefont {Y.~X.}\ \bibnamefont
  {Zhao}},\ and\ \bibinfo {author} {\bibfnamefont {S.~A.}\ \bibnamefont
  {Yang}},\ }\bibfield  {title} {\bibinfo {title} {Nodal surface semimetals:
  Theory and material realization},\ }\href
  {https://doi.org/10.1103/PhysRevB.97.115125} {\bibfield  {journal} {\bibinfo
  {journal} {Phys. Rev. B}\ }\textbf {\bibinfo {volume} {97}},\ \bibinfo
  {pages} {115125} (\bibinfo {year} {2018})}\BibitemShut {NoStop}%
\bibitem [{\citenamefont {Fang}\ \emph {et~al.}(2016)\citenamefont {Fang},
  \citenamefont {Weng}, \citenamefont {Dai},\ and\ \citenamefont
  {Fang}}]{ChenFangChinese2016}%
  \BibitemOpen
  \bibfield  {author} {\bibinfo {author} {\bibfnamefont {C.}~\bibnamefont
  {Fang}}, \bibinfo {author} {\bibfnamefont {H.}~\bibnamefont {Weng}}, \bibinfo
  {author} {\bibfnamefont {X.}~\bibnamefont {Dai}},\ and\ \bibinfo {author}
  {\bibfnamefont {Z.}~\bibnamefont {Fang}},\ }\bibfield  {title} {\bibinfo
  {title} {Topological nodal line semimetals},\ }\href
  {https://doi.org/10.1088/1674-1056/25/11/117106} {\bibfield  {journal}
  {\bibinfo  {journal} {Chinese Physics B}\ }\textbf {\bibinfo {volume} {25}},\
  \bibinfo {pages} {117106} (\bibinfo {year} {2016})}\BibitemShut {NoStop}%
\bibitem [{\citenamefont {Bzdu{\v{s}}ek}\ \emph {et~al.}(2016)\citenamefont
  {Bzdu{\v{s}}ek}, \citenamefont {Wu}, \citenamefont {R{\"u}egg}, \citenamefont
  {Sigrist},\ and\ \citenamefont {Soluyanov}}]{bzduvsek2016nodal}%
  \BibitemOpen
  \bibfield  {author} {\bibinfo {author} {\bibfnamefont {T.}~\bibnamefont
  {Bzdu{\v{s}}ek}}, \bibinfo {author} {\bibfnamefont {Q.}~\bibnamefont {Wu}},
  \bibinfo {author} {\bibfnamefont {A.}~\bibnamefont {R{\"u}egg}}, \bibinfo
  {author} {\bibfnamefont {M.}~\bibnamefont {Sigrist}},\ and\ \bibinfo {author}
  {\bibfnamefont {A.~A.}\ \bibnamefont {Soluyanov}},\ }\bibfield  {title}
  {\bibinfo {title} {Nodal-chain metals},\ }\href@noop {} {\bibfield  {journal}
  {\bibinfo  {journal} {Nature}\ }\textbf {\bibinfo {volume} {538}},\ \bibinfo
  {pages} {75} (\bibinfo {year} {2016})}\BibitemShut {NoStop}%
\bibitem [{\citenamefont {Liang}\ \emph {et~al.}(2016)\citenamefont {Liang},
  \citenamefont {Zhou}, \citenamefont {Yu}, \citenamefont {Wang},\ and\
  \citenamefont {Weng}}]{QiFengLiangPRB2016}%
  \BibitemOpen
  \bibfield  {author} {\bibinfo {author} {\bibfnamefont {Q.-F.}\ \bibnamefont
  {Liang}}, \bibinfo {author} {\bibfnamefont {J.}~\bibnamefont {Zhou}},
  \bibinfo {author} {\bibfnamefont {R.}~\bibnamefont {Yu}}, \bibinfo {author}
  {\bibfnamefont {Z.}~\bibnamefont {Wang}},\ and\ \bibinfo {author}
  {\bibfnamefont {H.}~\bibnamefont {Weng}},\ }\bibfield  {title} {\bibinfo
  {title} {Node-surface and node-line fermions from nonsymmorphic lattice
  symmetries},\ }\href {https://doi.org/10.1103/PhysRevB.93.085427} {\bibfield
  {journal} {\bibinfo  {journal} {Phys. Rev. B}\ }\textbf {\bibinfo {volume}
  {93}},\ \bibinfo {pages} {085427} (\bibinfo {year} {2016})}\BibitemShut
  {NoStop}%
\bibitem [{\citenamefont {Xie}\ \emph {et~al.}()\citenamefont {Xie},
  \citenamefont {Gao}, \citenamefont {Xu}, \citenamefont {Zhang}, \citenamefont
  {Hu},\ and\ \citenamefont {Law}}]{YingMingXieArxiv2020}%
  \BibitemOpen
  \bibfield  {author} {\bibinfo {author} {\bibfnamefont {Y.-M.}\ \bibnamefont
  {Xie}}, \bibinfo {author} {\bibfnamefont {X.-J.}\ \bibnamefont {Gao}},
  \bibinfo {author} {\bibfnamefont {X.~Y.}\ \bibnamefont {Xu}}, \bibinfo
  {author} {\bibfnamefont {C.-P.}\ \bibnamefont {Zhang}}, \bibinfo {author}
  {\bibfnamefont {J.-X.}\ \bibnamefont {Hu}},\ and\ \bibinfo {author}
  {\bibfnamefont {K.~T.}\ \bibnamefont {Law}},\ }\bibfield  {title} {\bibinfo
  {title} {Kramers nodal line metals},\ }\bibinfo {note} {arXiv:2008.03967
  (unpublished)}\BibitemShut {NoStop}%
\bibitem [{\citenamefont {Kavokin}(2004)}]{Kavokin}%
  \BibitemOpen
  \bibfield  {author} {\bibinfo {author} {\bibfnamefont {K.~V.}\ \bibnamefont
  {Kavokin}},\ }\bibfield  {title} {\bibinfo {title} {Symmetry of anisotropic
  exchange interactions in semiconductor nanostructures},\ }\href
  {https://doi.org/10.1103/PhysRevB.69.075302} {\bibfield  {journal} {\bibinfo
  {journal} {Phys. Rev. B}\ }\textbf {\bibinfo {volume} {69}},\ \bibinfo
  {pages} {075302} (\bibinfo {year} {2004})}\BibitemShut {NoStop}%
\bibitem [{\citenamefont {Kato}\ \emph {et~al.}(2003)\citenamefont {Kato},
  \citenamefont {Myers}, \citenamefont {Driscoll}, \citenamefont {Gossard},
  \citenamefont {Levy},\ and\ \citenamefont
  {Awschalom}}]{Kato-gfactorresonance}%
  \BibitemOpen
  \bibfield  {author} {\bibinfo {author} {\bibfnamefont {Y.}~\bibnamefont
  {Kato}}, \bibinfo {author} {\bibfnamefont {R.~C.}\ \bibnamefont {Myers}},
  \bibinfo {author} {\bibfnamefont {D.~C.}\ \bibnamefont {Driscoll}}, \bibinfo
  {author} {\bibfnamefont {A.~C.}\ \bibnamefont {Gossard}}, \bibinfo {author}
  {\bibfnamefont {J.}~\bibnamefont {Levy}},\ and\ \bibinfo {author}
  {\bibfnamefont {D.~D.}\ \bibnamefont {Awschalom}},\ }\bibfield  {title}
  {\bibinfo {title} {Gigahertz electron spin manipulation using
  voltage-controlled g-tensor modulation},\ }\href
  {https://doi.org/10.1126/science.1080880} {\bibfield  {journal} {\bibinfo
  {journal} {Science}\ }\textbf {\bibinfo {volume} {299}},\ \bibinfo {pages}
  {1201} (\bibinfo {year} {2003})}\BibitemShut {NoStop}%
\bibitem [{\citenamefont {Veldhorst}\ \emph {et~al.}(2015)\citenamefont
  {Veldhorst}, \citenamefont {Ruskov}, \citenamefont {Yang}, \citenamefont
  {Hwang}, \citenamefont {Hudson}, \citenamefont {Flatt\'e}, \citenamefont
  {Tahan}, \citenamefont {Itoh}, \citenamefont {Morello},\ and\ \citenamefont
  {Dzurak}}]{VeldhorstPRB}%
  \BibitemOpen
  \bibfield  {author} {\bibinfo {author} {\bibfnamefont {M.}~\bibnamefont
  {Veldhorst}}, \bibinfo {author} {\bibfnamefont {R.}~\bibnamefont {Ruskov}},
  \bibinfo {author} {\bibfnamefont {C.~H.}\ \bibnamefont {Yang}}, \bibinfo
  {author} {\bibfnamefont {J.~C.~C.}\ \bibnamefont {Hwang}}, \bibinfo {author}
  {\bibfnamefont {F.~E.}\ \bibnamefont {Hudson}}, \bibinfo {author}
  {\bibfnamefont {M.~E.}\ \bibnamefont {Flatt\'e}}, \bibinfo {author}
  {\bibfnamefont {C.}~\bibnamefont {Tahan}}, \bibinfo {author} {\bibfnamefont
  {K.~M.}\ \bibnamefont {Itoh}}, \bibinfo {author} {\bibfnamefont
  {A.}~\bibnamefont {Morello}},\ and\ \bibinfo {author} {\bibfnamefont {A.~S.}\
  \bibnamefont {Dzurak}},\ }\bibfield  {title} {\bibinfo {title} {Spin-orbit
  coupling and operation of multivalley spin qubits},\ }\href
  {https://doi.org/10.1103/PhysRevB.92.201401} {\bibfield  {journal} {\bibinfo
  {journal} {Phys. Rev. B}\ }\textbf {\bibinfo {volume} {92}},\ \bibinfo
  {pages} {201401} (\bibinfo {year} {2015})}\BibitemShut {NoStop}%
\bibitem [{\citenamefont {Crippa}\ \emph {et~al.}(2018)\citenamefont {Crippa},
  \citenamefont {Maurand}, \citenamefont {Bourdet}, \citenamefont
  {Kotekar-Patil}, \citenamefont {Amisse}, \citenamefont {Jehl}, \citenamefont
  {Sanquer}, \citenamefont {Lavi\'eville}, \citenamefont {Bohuslavskyi},
  \citenamefont {Hutin}, \citenamefont {Barraud}, \citenamefont {Vinet},
  \citenamefont {Niquet},\ and\ \citenamefont {De~Franceschi}}]{Crippa}%
  \BibitemOpen
  \bibfield  {author} {\bibinfo {author} {\bibfnamefont {A.}~\bibnamefont
  {Crippa}}, \bibinfo {author} {\bibfnamefont {R.}~\bibnamefont {Maurand}},
  \bibinfo {author} {\bibfnamefont {L.}~\bibnamefont {Bourdet}}, \bibinfo
  {author} {\bibfnamefont {D.}~\bibnamefont {Kotekar-Patil}}, \bibinfo {author}
  {\bibfnamefont {A.}~\bibnamefont {Amisse}}, \bibinfo {author} {\bibfnamefont
  {X.}~\bibnamefont {Jehl}}, \bibinfo {author} {\bibfnamefont {M.}~\bibnamefont
  {Sanquer}}, \bibinfo {author} {\bibfnamefont {R.}~\bibnamefont
  {Lavi\'eville}}, \bibinfo {author} {\bibfnamefont {H.}~\bibnamefont
  {Bohuslavskyi}}, \bibinfo {author} {\bibfnamefont {L.}~\bibnamefont {Hutin}},
  \bibinfo {author} {\bibfnamefont {S.}~\bibnamefont {Barraud}}, \bibinfo
  {author} {\bibfnamefont {M.}~\bibnamefont {Vinet}}, \bibinfo {author}
  {\bibfnamefont {Y.-M.}\ \bibnamefont {Niquet}},\ and\ \bibinfo {author}
  {\bibfnamefont {S.}~\bibnamefont {De~Franceschi}},\ }\bibfield  {title}
  {\bibinfo {title} {Electrical spin driving by $g$-matrix modulation in
  spin-orbit qubits},\ }\href {https://doi.org/10.1103/PhysRevLett.120.137702}
  {\bibfield  {journal} {\bibinfo  {journal} {Phys. Rev. Lett.}\ }\textbf
  {\bibinfo {volume} {120}},\ \bibinfo {pages} {137702} (\bibinfo {year}
  {2018})}\BibitemShut {NoStop}%
\bibitem [{\citenamefont {Schroer}\ \emph {et~al.}(2011)\citenamefont
  {Schroer}, \citenamefont {Petersson}, \citenamefont {Jung},\ and\
  \citenamefont {Petta}}]{Schroer-gfactor}%
  \BibitemOpen
  \bibfield  {author} {\bibinfo {author} {\bibfnamefont {M.~D.}\ \bibnamefont
  {Schroer}}, \bibinfo {author} {\bibfnamefont {K.~D.}\ \bibnamefont
  {Petersson}}, \bibinfo {author} {\bibfnamefont {M.}~\bibnamefont {Jung}},\
  and\ \bibinfo {author} {\bibfnamefont {J.~R.}\ \bibnamefont {Petta}},\
  }\bibfield  {title} {\bibinfo {title} {Field tuning the $g$ factor in inas
  nanowire double quantum dots},\ }\href
  {https://doi.org/10.1103/PhysRevLett.107.176811} {\bibfield  {journal}
  {\bibinfo  {journal} {Phys. Rev. Lett.}\ }\textbf {\bibinfo {volume} {107}},\
  \bibinfo {pages} {176811} (\bibinfo {year} {2011})}\BibitemShut {NoStop}%
\bibitem [{\citenamefont {Liles}\ \emph {et~al.}()\citenamefont {Liles},
  \citenamefont {Martins}, \citenamefont {Miserev}, \citenamefont {Kiselev},
  \citenamefont {Thorvaldson}, \citenamefont {Rendell}, \citenamefont {Jin},
  \citenamefont {Hudson}, \citenamefont {Veldhorst}, \citenamefont {Itoh},
  \citenamefont {Sushkov}, \citenamefont {Ladd}, \citenamefont {Dzurak},\ and\
  \citenamefont {Hamilton}}]{LilesArxiv2020}%
  \BibitemOpen
  \bibfield  {author} {\bibinfo {author} {\bibfnamefont {S.~D.}\ \bibnamefont
  {Liles}}, \bibinfo {author} {\bibfnamefont {F.}~\bibnamefont {Martins}},
  \bibinfo {author} {\bibfnamefont {D.~S.}\ \bibnamefont {Miserev}}, \bibinfo
  {author} {\bibfnamefont {A.~A.}\ \bibnamefont {Kiselev}}, \bibinfo {author}
  {\bibfnamefont {I.~D.}\ \bibnamefont {Thorvaldson}}, \bibinfo {author}
  {\bibfnamefont {M.~J.}\ \bibnamefont {Rendell}}, \bibinfo {author}
  {\bibfnamefont {I.~K.}\ \bibnamefont {Jin}}, \bibinfo {author} {\bibfnamefont
  {F.~E.}\ \bibnamefont {Hudson}}, \bibinfo {author} {\bibfnamefont
  {M.}~\bibnamefont {Veldhorst}}, \bibinfo {author} {\bibfnamefont {K.~M.}\
  \bibnamefont {Itoh}}, \bibinfo {author} {\bibfnamefont {O.~P.}\ \bibnamefont
  {Sushkov}}, \bibinfo {author} {\bibfnamefont {T.~D.}\ \bibnamefont {Ladd}},
  \bibinfo {author} {\bibfnamefont {A.~S.}\ \bibnamefont {Dzurak}},\ and\
  \bibinfo {author} {\bibfnamefont {A.~R.}\ \bibnamefont {Hamilton}},\
  }\bibinfo {title} {Electrical control of the g-tensor of a single hole in a
  silicon {MOS} quantum dot, {arXiv}:2012.04985 (unpublished)}\BibitemShut
  {NoStop}%
\bibitem [{\citenamefont {Souza}\ \emph {et~al.}(2016)\citenamefont {Souza},
  \citenamefont {Tomka}, \citenamefont {Kolodrubetz}, \citenamefont
  {Rosenberg},\ and\ \citenamefont {Polkovnikov}}]{Souza}%
  \BibitemOpen
\bibfield  {title} {  }\bibfield  {author} {\bibinfo {author} {\bibfnamefont
  {T.}~\bibnamefont {Souza}}, \bibinfo {author} {\bibfnamefont
  {M.}~\bibnamefont {Tomka}}, \bibinfo {author} {\bibfnamefont
  {M.}~\bibnamefont {Kolodrubetz}}, \bibinfo {author} {\bibfnamefont
  {S.}~\bibnamefont {Rosenberg}},\ and\ \bibinfo {author} {\bibfnamefont
  {A.}~\bibnamefont {Polkovnikov}},\ }\bibfield  {title} {\bibinfo {title}
  {Enabling adiabatic passages between disjoint regions in parameter space
  through topological transitions},\ }\href
  {https://doi.org/10.1103/PhysRevB.94.094106} {\bibfield  {journal} {\bibinfo
  {journal} {Phys. Rev. B}\ }\textbf {\bibinfo {volume} {94}},\ \bibinfo
  {pages} {094106} (\bibinfo {year} {2016})}\BibitemShut {NoStop}%
\bibitem [{\citenamefont {Curtright}\ \emph {et~al.}(2020)\citenamefont
  {Curtright}, \citenamefont {Cao}, \citenamefont {Huang}, \citenamefont
  {Sarmiento}, \citenamefont {Subedi}, \citenamefont {Tarrence},\ and\
  \citenamefont {Thapaliya}}]{Curtright2020charge}%
  \BibitemOpen
  \bibfield  {author} {\bibinfo {author} {\bibfnamefont {T.}~\bibnamefont
  {Curtright}}, \bibinfo {author} {\bibfnamefont {Z.}~\bibnamefont {Cao}},
  \bibinfo {author} {\bibfnamefont {S.}~\bibnamefont {Huang}}, \bibinfo
  {author} {\bibfnamefont {J.}~\bibnamefont {Sarmiento}}, \bibinfo {author}
  {\bibfnamefont {S.}~\bibnamefont {Subedi}}, \bibinfo {author} {\bibfnamefont
  {D.}~\bibnamefont {Tarrence}},\ and\ \bibinfo {author} {\bibfnamefont
  {T.}~\bibnamefont {Thapaliya}},\ }\bibfield  {title} {\bibinfo {title}
  {Charge densities for conducting ellipsoids},\ }\href@noop {} {\bibfield
  {journal} {\bibinfo  {journal} {European Journal of Physics}\ }\textbf
  {\bibinfo {volume} {41}},\ \bibinfo {pages} {035204} (\bibinfo {year}
  {2020})}\BibitemShut {NoStop}%
\bibitem [{\citenamefont {Sun}\ \emph {et~al.}(2018)\citenamefont {Sun},
  \citenamefont {Zhang},\ and\ \citenamefont {Bzdusek}}]{SunPRL2018}%
  \BibitemOpen
  \bibfield  {author} {\bibinfo {author} {\bibfnamefont {X.-Q.}\ \bibnamefont
  {Sun}}, \bibinfo {author} {\bibfnamefont {S.-C.}\ \bibnamefont {Zhang}},\
  and\ \bibinfo {author} {\bibfnamefont {T.}~\bibnamefont {Bzdusek}},\
  }\bibfield  {title} {\bibinfo {title} {Conversion rules for weyl points and
  nodal lines in topological media},\ }\href
  {https://doi.org/10.1103/PhysRevLett.121.106402} {\bibfield  {journal}
  {\bibinfo  {journal} {Phys. Rev. Lett.}\ }\textbf {\bibinfo {volume} {121}},\
  \bibinfo {pages} {106402} (\bibinfo {year} {2018})}\BibitemShut {NoStop}%
\bibitem [{\citenamefont {Schroer}\ \emph {et~al.}(2014)\citenamefont
  {Schroer}, \citenamefont {Kolodrubetz}, \citenamefont {Kindel}, \citenamefont
  {Sandberg}, \citenamefont {Gao}, \citenamefont {Vissers}, \citenamefont
  {Pappas}, \citenamefont {Polkovnikov},\ and\ \citenamefont
  {Lehnert}}]{Schroer_chern}%
  \BibitemOpen
  \bibfield  {author} {\bibinfo {author} {\bibfnamefont {M.~D.}\ \bibnamefont
  {Schroer}}, \bibinfo {author} {\bibfnamefont {M.~H.}\ \bibnamefont
  {Kolodrubetz}}, \bibinfo {author} {\bibfnamefont {W.~F.}\ \bibnamefont
  {Kindel}}, \bibinfo {author} {\bibfnamefont {M.}~\bibnamefont {Sandberg}},
  \bibinfo {author} {\bibfnamefont {J.}~\bibnamefont {Gao}}, \bibinfo {author}
  {\bibfnamefont {M.~R.}\ \bibnamefont {Vissers}}, \bibinfo {author}
  {\bibfnamefont {D.~P.}\ \bibnamefont {Pappas}}, \bibinfo {author}
  {\bibfnamefont {A.}~\bibnamefont {Polkovnikov}},\ and\ \bibinfo {author}
  {\bibfnamefont {K.~W.}\ \bibnamefont {Lehnert}},\ }\bibfield  {title}
  {\bibinfo {title} {Measuring a topological transition in an artificial
  spin-$1/2$ system},\ }\href {https://doi.org/10.1103/PhysRevLett.113.050402}
  {\bibfield  {journal} {\bibinfo  {journal} {Phys. Rev. Lett.}\ }\textbf
  {\bibinfo {volume} {113}},\ \bibinfo {pages} {050402} (\bibinfo {year}
  {2014})}\BibitemShut {NoStop}%
\bibitem [{\citenamefont {Roushan}\ \emph {et~al.}(2014)\citenamefont
  {Roushan}, \citenamefont {Neill}, \citenamefont {Chen}, \citenamefont
  {Kolodrubetz}, \citenamefont {Quintana}, \citenamefont {Leung}, \citenamefont
  {Fang}, \citenamefont {Barends}, \citenamefont {Campbell}, \citenamefont
  {Chen}, \citenamefont {Chiaro}, \citenamefont {Dunsworth}, \citenamefont
  {Jeffrey}, \citenamefont {Kelly}, \citenamefont {Megrant}, \citenamefont
  {Mutus}, \citenamefont {O'Malley}, \citenamefont {Sank}, \citenamefont
  {Vainsencher}, \citenamefont {Wenner}, \citenamefont {White}, \citenamefont
  {Polkovnikov}, \citenamefont {Cleland},\ and\ \citenamefont
  {Martinis}}]{Roushan}%
  \BibitemOpen
  \bibfield  {author} {\bibinfo {author} {\bibfnamefont {P.}~\bibnamefont
  {Roushan}}, \bibinfo {author} {\bibfnamefont {C.}~\bibnamefont {Neill}},
  \bibinfo {author} {\bibfnamefont {Y.}~\bibnamefont {Chen}}, \bibinfo {author}
  {\bibfnamefont {M.}~\bibnamefont {Kolodrubetz}}, \bibinfo {author}
  {\bibfnamefont {C.}~\bibnamefont {Quintana}}, \bibinfo {author}
  {\bibfnamefont {N.}~\bibnamefont {Leung}}, \bibinfo {author} {\bibfnamefont
  {M.}~\bibnamefont {Fang}}, \bibinfo {author} {\bibfnamefont {R.}~\bibnamefont
  {Barends}}, \bibinfo {author} {\bibfnamefont {B.}~\bibnamefont {Campbell}},
  \bibinfo {author} {\bibfnamefont {Z.}~\bibnamefont {Chen}}, \bibinfo {author}
  {\bibfnamefont {B.}~\bibnamefont {Chiaro}}, \bibinfo {author} {\bibfnamefont
  {A.}~\bibnamefont {Dunsworth}}, \bibinfo {author} {\bibfnamefont
  {E.}~\bibnamefont {Jeffrey}}, \bibinfo {author} {\bibfnamefont
  {J.}~\bibnamefont {Kelly}}, \bibinfo {author} {\bibfnamefont
  {A.}~\bibnamefont {Megrant}}, \bibinfo {author} {\bibfnamefont
  {J.}~\bibnamefont {Mutus}}, \bibinfo {author} {\bibfnamefont {P.~J.~J.}\
  \bibnamefont {O'Malley}}, \bibinfo {author} {\bibfnamefont {D.}~\bibnamefont
  {Sank}}, \bibinfo {author} {\bibfnamefont {A.}~\bibnamefont {Vainsencher}},
  \bibinfo {author} {\bibfnamefont {J.}~\bibnamefont {Wenner}}, \bibinfo
  {author} {\bibfnamefont {T.}~\bibnamefont {White}}, \bibinfo {author}
  {\bibfnamefont {A.}~\bibnamefont {Polkovnikov}}, \bibinfo {author}
  {\bibfnamefont {A.~N.}\ \bibnamefont {Cleland}},\ and\ \bibinfo {author}
  {\bibfnamefont {J.~M.}\ \bibnamefont {Martinis}},\ }\bibfield  {title}
  {\bibinfo {title} {Observation of topological transitions in interacting
  quantum circuits},\ }\href@noop {} {\bibfield  {journal} {\bibinfo  {journal}
  {Nature}\ }\textbf {\bibinfo {volume} {515}},\ \bibinfo {pages} {241}
  (\bibinfo {year} {2014})}\BibitemShut {NoStop}%
\bibitem [{\citenamefont {Koppens}\ \emph {et~al.}(2006)\citenamefont
  {Koppens}, \citenamefont {Buizert}, \citenamefont {Tielrooij}, \citenamefont
  {Vink}, \citenamefont {Nowack}, \citenamefont {Meunier}, \citenamefont
  {Kouwenhoven},\ and\ \citenamefont {Vandersypen}}]{Koppens-esr}%
  \BibitemOpen
  \bibfield  {author} {\bibinfo {author} {\bibfnamefont {F.~H.~L.}\
  \bibnamefont {Koppens}}, \bibinfo {author} {\bibfnamefont {C.}~\bibnamefont
  {Buizert}}, \bibinfo {author} {\bibfnamefont {K.~J.}\ \bibnamefont
  {Tielrooij}}, \bibinfo {author} {\bibfnamefont {I.~T.}\ \bibnamefont {Vink}},
  \bibinfo {author} {\bibfnamefont {K.~C.}\ \bibnamefont {Nowack}}, \bibinfo
  {author} {\bibfnamefont {T.}~\bibnamefont {Meunier}}, \bibinfo {author}
  {\bibfnamefont {L.~P.}\ \bibnamefont {Kouwenhoven}},\ and\ \bibinfo {author}
  {\bibfnamefont {L.~M.~K.}\ \bibnamefont {Vandersypen}},\ }\bibfield  {title}
  {\bibinfo {title} {Driven coherent oscillations of a single electron spin in
  a quantum dot},\ }\href {https://doi.org/10.1038/nature05065} {\bibfield
  {journal} {\bibinfo  {journal} {Nature}\ }\textbf {\bibinfo {volume} {442}},\
  \bibinfo {pages} {766} (\bibinfo {year} {2006})}\BibitemShut {NoStop}%
\bibitem [{\citenamefont {Petta}\ \emph {et~al.}(2010)\citenamefont {Petta},
  \citenamefont {Lu},\ and\ \citenamefont {Gossard}}]{Petta_beamsplitter}%
  \BibitemOpen
  \bibfield  {author} {\bibinfo {author} {\bibfnamefont {J.~R.}\ \bibnamefont
  {Petta}}, \bibinfo {author} {\bibfnamefont {H.}~\bibnamefont {Lu}},\ and\
  \bibinfo {author} {\bibfnamefont {A.~C.}\ \bibnamefont {Gossard}},\
  }\bibfield  {title} {\bibinfo {title} {A coherent beam splitter for
  electronic spin states},\ }\href@noop {} {\bibfield  {journal} {\bibinfo
  {journal} {Science}\ }\textbf {\bibinfo {volume} {327}},\ \bibinfo {pages}
  {669} (\bibinfo {year} {2010})}\BibitemShut {NoStop}%
\bibitem [{\citenamefont {Tanttu}\ \emph {et~al.}(2019)\citenamefont {Tanttu},
  \citenamefont {Hensen}, \citenamefont {Chan}, \citenamefont {Yang},
  \citenamefont {Huang}, \citenamefont {Fogarty}, \citenamefont {Hudson},
  \citenamefont {Itoh}, \citenamefont {Culcer}, \citenamefont {Laucht},
  \citenamefont {Morello},\ and\ \citenamefont {Dzurak}}]{Tanttu}%
  \BibitemOpen
  \bibfield  {author} {\bibinfo {author} {\bibfnamefont {T.}~\bibnamefont
  {Tanttu}}, \bibinfo {author} {\bibfnamefont {B.}~\bibnamefont {Hensen}},
  \bibinfo {author} {\bibfnamefont {K.~W.}\ \bibnamefont {Chan}}, \bibinfo
  {author} {\bibfnamefont {C.~H.}\ \bibnamefont {Yang}}, \bibinfo {author}
  {\bibfnamefont {W.~W.}\ \bibnamefont {Huang}}, \bibinfo {author}
  {\bibfnamefont {M.}~\bibnamefont {Fogarty}}, \bibinfo {author} {\bibfnamefont
  {F.}~\bibnamefont {Hudson}}, \bibinfo {author} {\bibfnamefont
  {K.}~\bibnamefont {Itoh}}, \bibinfo {author} {\bibfnamefont {D.}~\bibnamefont
  {Culcer}}, \bibinfo {author} {\bibfnamefont {A.}~\bibnamefont {Laucht}},
  \bibinfo {author} {\bibfnamefont {A.}~\bibnamefont {Morello}},\ and\ \bibinfo
  {author} {\bibfnamefont {A.}~\bibnamefont {Dzurak}},\ }\bibfield  {title}
  {\bibinfo {title} {Controlling spin-orbit interactions in silicon quantum
  dots using magnetic field direction},\ }\href
  {https://doi.org/10.1103/PhysRevX.9.021028} {\bibfield  {journal} {\bibinfo
  {journal} {Phys. Rev. X}\ }\textbf {\bibinfo {volume} {9}},\ \bibinfo {pages}
  {021028} (\bibinfo {year} {2019})}\BibitemShut {NoStop}%
\bibitem [{\citenamefont {Garg}(2010)}]{Garg}%
  \BibitemOpen
  \bibfield  {author} {\bibinfo {author} {\bibfnamefont {A.}~\bibnamefont
  {Garg}},\ }\bibfield  {title} {\bibinfo {title} {{Berry} phases near
  degeneracies: Beyond the simplest case},\ }\href
  {https://doi.org/10.1119/1.3377135} {\bibfield  {journal} {\bibinfo
  {journal} {Am. J. Phys.}\ }\textbf {\bibinfo {volume} {78}},\ \bibinfo
  {pages} {661} (\bibinfo {year} {2010})}\BibitemShut {NoStop}%
\bibitem [{\citenamefont {Wiesendanger}(2009)}]{Wiesendanger}%
  \BibitemOpen
  \bibfield  {author} {\bibinfo {author} {\bibfnamefont {R.}~\bibnamefont
  {Wiesendanger}},\ }\bibfield  {title} {\bibinfo {title} {Spin mapping at the
  nanoscale and atomic scale},\ }\href
  {https://doi.org/10.1103/RevModPhys.81.1495} {\bibfield  {journal} {\bibinfo
  {journal} {Rev. Mod. Phys.}\ }\textbf {\bibinfo {volume} {81}},\ \bibinfo
  {pages} {1495} (\bibinfo {year} {2009})}\BibitemShut {NoStop}%
\bibitem [{\citenamefont {Spinelli}\ \emph {et~al.}(2015)\citenamefont
  {Spinelli}, \citenamefont {Gerrits}, \citenamefont {Toskovic}, \citenamefont
  {Bryant}, \citenamefont {Ternes},\ and\ \citenamefont {Otte}}]{Spinelli}%
  \BibitemOpen
  \bibfield  {author} {\bibinfo {author} {\bibfnamefont {A.}~\bibnamefont
  {Spinelli}}, \bibinfo {author} {\bibfnamefont {M.}~\bibnamefont {Gerrits}},
  \bibinfo {author} {\bibfnamefont {R.}~\bibnamefont {Toskovic}}, \bibinfo
  {author} {\bibfnamefont {B.}~\bibnamefont {Bryant}}, \bibinfo {author}
  {\bibfnamefont {M.}~\bibnamefont {Ternes}},\ and\ \bibinfo {author}
  {\bibfnamefont {A.~F.}\ \bibnamefont {Otte}},\ }\bibfield  {title} {\bibinfo
  {title} {Exploring the phase diagram of the two-impurity {Kondo} problem},\
  }\href@noop {} {\bibfield  {journal} {\bibinfo  {journal} {Nature
  Communications}\ }\textbf {\bibinfo {volume} {6}},\ \bibinfo {pages} {10046}
  (\bibinfo {year} {2015})}\BibitemShut {NoStop}%
\bibitem [{\citenamefont {Horn}\ and\ \citenamefont
  {Johnson}(1985)}]{horn1985}%
  \BibitemOpen
  \bibfield  {author} {\bibinfo {author} {\bibfnamefont {R.~A.}\ \bibnamefont
  {Horn}}\ and\ \bibinfo {author} {\bibfnamefont {C.~R.}\ \bibnamefont
  {Johnson}},\ }\href@noop {} {\emph {\bibinfo {title} {Matrix Analysis}}}\
  (\bibinfo  {publisher} {Cambridge, MA: Cambridge University Press},\ \bibinfo
  {year} {1985})\BibitemShut {NoStop}%
\end{thebibliography}%

\newpage
\appendix

\begin{widetext}

\section{Proofs of Eq.~\eqref{eq:electrostatics}}
\label{app:electrostatics}

\subsection{First proof}
\label{app:firstproof}

In the main text, Eq.~\eqref{eq:electrostatics} expresses
the linear charge density $\nu(s)$ of a charged
loop in terms of the electric field $\vec E(\vec r)$ 
created by the loop.
Here, we provide an elementary proof of that result.

Consider a cylindrical section of the torus,
together with its top base and bottom base, 
surrounding a section of the degeneracy circle
in Fig.~\ref{fig:torus}~a.
Without loss of generality, we can take the section 
defined by the interval $s\in [0,s_0]$ with 
$0 < s_0 < 2\pi R$.
From Gauss' law, the total charge enclosed by the 
cylinder is expressed from the electric field as
\begin{align}
\frac{1}{\varepsilon_0} \int_0^{s_0} ds \nu(s) 
& = \int_\text{cylinder} d \vec{A} \cdot \vec E.  \\
\intertext{Up to now, we assume that the cylinder has a nonzero meridian radius $r$. Splitting up the cylinder's surface integral to its three parts, we obtain}
\frac{1}{\varepsilon_0} \int_0^{s_0} ds \nu(s) 
& = \int_{\text{top}\, \text{base}} d \vec{A} \cdot \vec E
+ \int_{\text{bottom}\, \text{base}} d \vec{A} \cdot \vec E
+  \int_{\text{side}} d \vec{A} \cdot \vec E.  \\
\intertext{For $r \to 0$, the top base and bottom base contributions converge to zero (see below), 
hence we find}
\frac{1}{\varepsilon_0} \int_0^{s_0} ds \nu(s) 
& = \lim_{r\to 0} \int_{\text{side}} d \vec{A} \cdot \vec E.  \\
\intertext{Using the parametrization $\vec p_r(s,\vartheta)$ of the main text, this can be written as}
\frac{1}{\varepsilon_0} \int_0^{s_0} ds \nu(s) 
& = 
\lim_{r\to 0}
\int_0^{s_0} ds \int_0^{2\pi} d \vartheta
\vec E (\vec p_r(s,\vartheta)) \cdot
\left[
 \partial_\vartheta \vec{p}_r \times \partial_s \vec{p}_r
\right]_{s,\vartheta}.
\end{align}
Assuming that the limit and the $s$-integral can be
exchanged, 
and considering that the boundaries of 
the $s$ integral were arbitrary, we arrive to Eq.~\eqref{eq:electrostatics}.

In the remaining part of this subsection, we prove that 
the flux contributions of the top and bottom bases approach zero as the radius $r$ 
of the cylindrical section of the torus also approaches zero.
To prove this, we consider a rather general setting depicted 
in Fig.~\ref{fig:appA}.
Here, the red curve is a charged wire, parametrized as 
$\vec x(s) = (x(s), y(s), z(s))$ by its path length $s$, 
such that its left end corresponds to $s=0$.
The linear charge density of the wire is $\nu(s)$. 
Our goal is to show that the electric flux piercing the 
disk (blue) of radius $r$ at $s=0$, chosen to be perpendicular to the wire, 
converges to zero as the radius $r$ approaches zero.
This in turn ensures that the top-base and bottom-base integrals in the preceding 
paragraph vanish in the limit $r \to 0$. 

\begin{figure}
	\begin{center}
		\includegraphics[width=0.5\columnwidth]{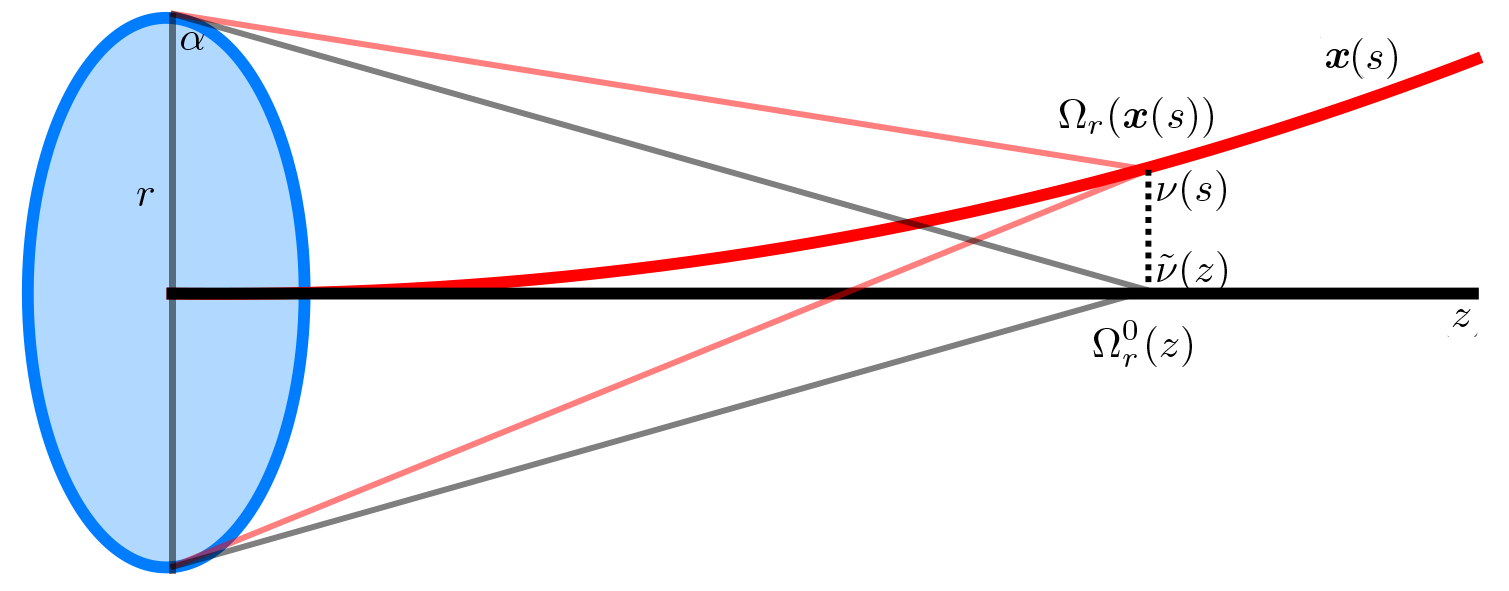}
	\end{center}
	\caption{{\bf Electric flux created by a charged wire (red) 
	through a disk (blue).}
	Section~\ref{app:firstproof} proves that the flux converges to 
	zero as the disk radius $r$ approaches zero.
	\label{fig:appA}}
\end{figure}
 
The flux created by a small line element of the charged wire between 
$\vec x(s)$ and $\vec x(s+ds)$ is $(4\pi \varepsilon_0)^{-1}\Omega_r(\vec x(s)) \nu(s) ds$,
where $\Omega_r(\vec x(s))$ is the solid angle under which the disk is seen
from the point $\vec x(s)$, see Fig.~\ref{fig:appA}.
Without loss of generality, we assume that $\nu(s)$ is positive.
Then, the flux created by the charged wire is expressed as:
\bean\label{eq:fluxclose}
\Phi_E=\int_\text{disk}\vec E\cdot d\vec A
=
\frac{1}{4\pi\varepsilon_0}\int\limits_{0}^{s_\text{max}}\Omega_r(\vec x(s))\nu(s)ds
=
\frac{1}{4\pi\varepsilon_0}\int\limits_{0}^{z_\text{max}}\Omega_r(\vec x(s(z)))\tilde{\nu}(z)dz.
\eean
Here, $s_\text{max}$ is the path length parameter value corresponding to the right end of the wire, 
$z_\text{max} = z(s_\text{max})$ is the corresponding $z$ coordinate, and
$s(z)$ is the inverse function of the parametrization component $z(s)$ which
is assumed to be invertible. 
In the last step of Eq.~\eqref{eq:fluxclose},
we substituted the integration variable $s$
by the $z$ coordinate, and accordingly, we introduced the modified linear charge density $\tilde{\nu}(z)=\nu(s(z))\cdot\frac{ds}{dz}=
\nu(s(z)) \sqrt{1+\left(\frac{dx}{dz}\right)^2+\left(\frac{dy}{dz}\right)^2}$,  
which specifies the charge on the wire per unit distance along the $z$ axis. 

Now we give an upper bound to the flux, by substituting the 
modified charge density by its maximum, $\tilde{\nu}_\text{max}$:
\bean\label{eq:fluxzaxis}
\Phi_E\le
\Phi_E^{\text{(1)}} = 
\frac{\tilde{\nu}_{\text{max}}}{4\pi\varepsilon_0}\int\limits_0^{z_\text{max}}\Omega_r(\vec x(s(z)))dz.
\eean
We give a further, looser upper bound to the flux, by utilizing the 
relation between the solid angles of the points of the wire and the points of the $z$
axis, namely, $\Omega_r(\vec x(s(z))) \le \Omega_r(0,0,z)$, see
Fig.~\ref{fig:appA}:
\bean\label{eq:fluxzaxis2}
\Phi_E\le
\Phi_E^{\text{(1)}} \le
\Phi_E^{\text{(2)}} = 
\frac{\tilde{\nu}_{\text{max}}}{4\pi\varepsilon_0}\int\limits_{0}^{z_\text{max}}\Omega_r(0,0,z)dz.
\eean
Using $\Omega_r(0,0,z) = 2\pi \left(1- \frac{z}{\sqrt{r^2+z^2}}\right)$,
and performing a second change of variable
by introducing the angle \mbox{$\alpha=\tan^{-1}\left(\frac{z}{r}\right)$},
see Fig.~\ref{fig:appA}, we find:
\bean\label{eq:fluxalpha}
\Phi_E^{(2)}
=\frac{\tilde{\nu}_{\text{max}}}{2\varepsilon_0}\int\limits_0^{\alpha_\text{max}}(1-\sin\alpha)\frac{rd\alpha}{\cos^2\alpha} =
\frac{\tilde{\nu}_{\text{max}}r}{2\varepsilon_0}\int\limits_0^{\alpha_\text{max}}\frac{d\alpha}{(1+\sin\alpha)}=\frac{\tilde{\nu}_{\text{max}}r}{\varepsilon_0}\frac{1}{\text{ctg}\left(\frac{\alpha_{\text{max}}}{2}\right)+1}.
\eean
As we decrease $r$ to zero, $\alpha_\text{max}$ converges to $\pi/2$, hence
the second fraction converges to $1/2$.
Therefore, this upper bound $\Phi_E^{(2)}$ converges to zero as $r\to 0$,
and hence the same is true for the flux $\Phi_E$.

\subsection{Second proof}

To give an alternative proof of the charge density 
formula \eqref{eq:electrostatics}, 
we start with the electrostatic field
created by a linear charge density:
\begin{equation}
    {\bm E}({\bm r}) = \frac{1}{4\pi \varepsilon_0} \int ds\; \nu(s) \frac{{\bm r} - {\bm r}(s)}{|{\bm r} - {\bm r}(s)|^3},
\end{equation}
where ${\bm r}(s) = (0,0,s)$ 
is the parametrization of the line charge.
For simplicity, we assume that 
the line charge is located along the 
$z$ axis.
We justify this assumption \emph{a posteriori} by the fact that the contribution of the charge distribution far from the point in question to the surface integral vanishes in the $r\to 0$ limit.
If the degeneracy line is smooth, in the $r\to 0$ limit it is locally well approximated by a straight line.

Substituting this into the integral for a cylinder surrounding the charge density:
\begin{equation}
\lim_{r\to 0}
\int_0^{2\pi} d \vartheta
\vec E (\vec p_r(s,\vartheta)) \cdot
\left[
 \partial_\vartheta \vec{p}_r \times \partial_s \vec{p}_r
\right]_{s,\vartheta} = 
\lim_{r\to 0}
\frac{1}{\varepsilon_0}
\int ds'\;\nu(s') \frac{1}{2r \left[1 + \left(\frac{s - s'}{r}\right)^2\right]^{\frac{3}{2}}} = \frac{\nu(s)}{\varepsilon_0}
\end{equation}
where $\vec{p}_r(\vartheta, s) = (r \cos \vartheta, r \sin\vartheta, s)$.
In the last step, we used that the fraction in the integrand converges to $\delta(s - s')$ in the $r\to 0$ limit.

\section{Numerical techniques to obtain the figures}
\label{sec:numerics}

In this Appendix, we outline the numerical 
techniques we used to obtain
Figs.~\ref{fig:neutral} and
\ref{fig:charged} of the main text.
\subsection{Berry flux density - Figs.~\ref{fig:neutral}~a and~\ref{fig:charged}~a}

Figs.~\ref{fig:neutral}~a and 
\ref{fig:charged}~a show the Berry flux density
$\mathcal{B}_n$, defined via
Eqs.~\eqref{eq:berryfluxdensity}
and \eqref{eq:componentwise},
on the surface of the torus
surrounding the circular degeneracy line.
To obtain Fig.~\ref{fig:neutral}~a,
our first step was to define
a $300 \times 300$ square grid
on the $s,\vartheta$ parameter space, 
and the corresponding grid of points on the torus,
obtained via the parametrization $\vec p_r(s,\vartheta)$.
The second step was to numerically approximate
the derivatives in Eq.~\eqref{eq:componentwise}
on the grid points of the torus
as 
\begin{equation}
    \ket{\partial_{B_\alpha}\psi_0(\vec B)}\approx \frac{\ket{\psi_0(\vec B+
    \delta \vec e_\alpha)}-\ket{\psi_0(\vec B- \delta\vec e_\alpha)}}{2 \delta}.
\end{equation}
Here, $\alpha \in \{x,y,z\}$, 
$\vec{e}_\alpha$ is the canonical 
unit vector pointing in direction $\alpha$,
and we used $\delta = 10^{-9}$.
The gauge of the ground-state wave functions in the vicinity
of a given $\vec B$ was fixed such that 
the greatest-magnitude component of the four-component wave function
was chosen to be real and positive.
Having the Berry curvature vector field
$\vec{\mathcal{B}}$ 
at hand, our final step was to 
evalute the Berry flux density by 
taking the normal projection in
Eq.~\eqref{eq:berryfluxdensity}.
The data shown in Fig.~\ref{fig:charged}~a is obtained
similarly.
\subsection{2D Berry curvature - Figs.~\ref{fig:neutral}~b and~\ref{fig:charged}~b}

The 2D Berry curvature
$\mathcal{B}_\text{2D}(s,\vartheta)$, plotted in 
Figs.~\ref{fig:neutral}~b and~\ref{fig:charged}~b,
is defined in Eq.~\eqref{eq:2dberrycurvature}
of the main text.
We claim that this 2D Berry curvature is
related to the Berry flux density discussed
in the previous section via
\begin{eqnarray}
\label{eq:bbrelation}
    \mathcal{B}_\text{2D} (s, \vartheta) &=&
    r\left(1+\frac{r}{R}\sin\vartheta\right)
    \mathcal{B}_n(\vec p_r(s,\vartheta)).
\end{eqnarray}
Using this relation, 
we converted the data in Fig.~\ref{fig:neutral}~a
(Fig.~\ref{fig:charged}~a)
to the data in Fig.~\ref{fig:neutral}~b
(Fig.~\ref{fig:charged}~b). 
The proof of Eq.~\eqref{eq:bbrelation}
is straightforward:
Eq.~\eqref{eq:normalvector} is
used at the right hand side of 
Eq.~\eqref{eq:2dberrycurvature},
then Eq.~\eqref{eq:berryfluxdensity}
is used, and finally the 
absolute value 
$|\partial_\vartheta \vec p_r(s,\vartheta) \times 
\partial_s \vec p_r(s,\vartheta) |$
is evaluated using the specific
parametrization in Eq.~\eqref{eq:parametrization}.

As a side remark, we also claim that 
the 2D Berry curvature can be expressed as
\begin{equation}
\label{eq:2dalternative}
    \mathcal{B}_\text{2D}(s,\vartheta)=
    -2\operatorname{Im}\braket{
    \partial_{\vartheta} \tilde{\psi}_0|
    \partial_{s} \tilde{\psi}_0},
\end{equation}
where $\tilde{\psi}_0(s,\vartheta) 
= (\psi_0 \circ \vec p_r)(s,\vartheta)$.
This is proven using the 
chain rule, which implies
\begin{eqnarray}
    \bra{\partial_\vartheta \tilde{\psi}_0}&=&\sum \limits_{\alpha\in\{x,y,z\}}(\partial_\vartheta p_{r,\alpha})\bra{\partial_{B_\alpha} \psi_0},\\
    \ket{\partial_s \tilde{\psi}_0}&=&\sum \limits_{\beta\in\{x,y,z\}}(\partial_s p_{r,\beta})\ket{\partial_{B_\beta} \psi_0}.
\end{eqnarray}
With these, we find
\begin{align}
    -2\operatorname{Im}\braket{
    \partial_{\vartheta} \tilde{\psi}_0|
    \partial_{s} \tilde{\psi}_0} &=\sum \limits_{\alpha,\beta\in\{x,y,z\}}-2\operatorname{Im}(\partial_\vartheta p_{r,\alpha})(\partial_s p_{r,\beta})\braket{\partial_{B_\alpha} \psi_0|\partial_{B_\beta} \psi_0}.  \\
\intertext{Using Eq.~\eqref{eq:componentwise}, the right hand side
is transformed as }
-2\operatorname{Im}\braket{
    \partial_{\vartheta} \tilde{\psi}_0|
    \partial_{s} \tilde{\psi}_0}
    &=\sum \limits_{\alpha,\beta\in\{x,y,z\}}(\partial_\vartheta p_{r,\alpha})(\partial_s p_{r,\beta})\epsilon_{\alpha\beta\gamma}\mathcal{B}_{\gamma}
    =[(\partial_\vartheta \vec p_r)\times (\partial_s \vec p_r)]\cdot \mathbfcal{B}.
\end{align}
This, together with Eq.~\eqref{eq:2dberrycurvature},
concludes the proof.

\subsection{Apparent linear topological charge density - Figs.~\ref{fig:neutral}~c and~\ref{fig:charged}~c}

The apparent linear topological charge density $\tilde{\nu}_r(s)$
is defined in Eq.~\eqref{eq:apparent}. 
To obtain the data shown in panels Fig.~\ref{fig:neutral}~c
[Fig.~\ref{fig:charged}~c], we performed a 
numerical $\vartheta$ integration
of $\mathcal{B}_\text{2D}(s,\vartheta)$,
using an $N \times N$ grid in $(s,\vartheta)$-space:
\begin{equation}
    \tilde{\nu}_r=\frac{1}{2\pi}\int_0^{2\pi} \mathcal{B}_\text{2D}(s,\vartheta) d\vartheta
    \approx\frac{1}{N}\sum\limits_{k=1}^N\mathcal{B}_\text{2D}\left(s,k\cdot\frac{2\pi}{N}\right),
\end{equation}
with $N = 300$ [$N=1000$].

\subsection{Chern number - Figs.~\ref{fig:neutral}~d and 
\ref{fig:charged}~d}

The ground-state Chern number $\mathcal{Q}$ on 
the torus, as a function of the meridian radius $r$,
is shown in 
Fig.~\ref{fig:neutral}~d and Fig.~\ref{fig:charged}~d.
This quantity is obtained from the apparent
topological charge density of panel c,
via a numerical integration over 
the longitude path length $s$, following
\begin{equation}
    \mathcal{Q}_r
    =
    \frac{1}{2\pi}\int_S d\vec A\cdot\mathbfcal{B}
    = \frac{1}{2\pi} \int_0^{2\pi R} \int_0^{2\pi}
    \mathcal{B}_\text{2D}(s,\vartheta) d \vartheta d s
    = 
     \int_0^{2\pi R} 
    \tilde{\nu}_r (s)  d s
    \approx\frac{2 \pi R}{N}\sum\limits_{k=1}^N\tilde{\nu}_r\left(k\cdot\frac{2\pi R}{N}\right).
\end{equation}

\section{Jordan normal forms for the
examples in the main text}
\label{sec:jordannormalforms}

In this section, we revisit the Jordan decomposition of
$3\times 3$ real matrices, and discuss the relation
between the matrix $\hat{\vec{M}}$ introduced in Eq.~\eqref{eq:matrixm} of the
main text, its Jordan decomposition,
and the magnetic degeneracy points.

In the main text, we have introduced the real valued non-symmetric matrix $\hat{\vec M}$ 
in Eq.~\eqref{eq:matrixm},
as the central quantity of the
two-spin problem.
Also, we claimed that 
the directions of the degeneracy points
are described by the left 
eigenvectors (left \textit{ordinary eigenvectors}) 
of this matrix $\hat{\vec M}$. 
Because of the non-symmetric property, 
$\hat{\vec M}$ is not always diagonalizable. 
Instead, it can be written as
\bean\label{eq:jordan}
\hat{\vec M} =\hat{\vec P}\hat{\vec J}\hat{\vec P}^{-1}=
\left[
  \begin{array}{cccc}
    \vertbar & \vertbar & \vertbar \\
    \vec w_{1}    & \vec w_{2}    & \vec w_{2}\\
    \vertbar & \vertbar & \vertbar 
  \end{array}
\right]
\hat{\vec J}
\left[
  \begin{array}{ccc}
    \horzbar & \vec v_{1}^{\text{T}} & \horzbar \\
    \horzbar & \vec v_{2}^{\text{T}} & \horzbar \\
    \horzbar & \vec v_{3}^{\text{T}} & \horzbar
  \end{array}
\right],
\eean
which is called the Jordan decomposition~\cite{horn1985}.
Here, $\hat{\vec P}^{-1}$ ($\hat{\vec P}$) is a non-singular matrix whose rows (columns) are the left (right) \textit{generalized eigenvectors} of $\hat{\vec M}$. From $\hat{\vec P}^{-1}\hat{\vec P}=\mathds{1}$ it follows that $\vec v_{i}^{\text{T}}\vec w_{j}=\delta_{ij}$. It is important to note that the left (right) generalized 
eigenvectors are not necessarily orthogonal to each other and not necessarily normalized. Also, the transformation matrix $\hat{\vec P}$ is not unique. The matrix $\hat{\vec J}$ is the Jordan normal form 
of $\hat{\vec M}$,
which has a block-diagonal structure formed of Jordan
blocks, matrices with the following structure:
\bean
\label{eq:jordanblock}
\hat{\vec J}_{1\times 1}(\lambda)=
\begin{pmatrix}
\lambda
\end{pmatrix},\hspace{6mm}
\hat{\vec J}_{2\times 2}(\lambda)=
\begin{pmatrix}
\lambda&1\\
0&\lambda
\end{pmatrix},\hspace{6mm}
\hat{\vec J}_{3\times 3}(\lambda)=
\begin{pmatrix}
\lambda&1&0\\
0&\lambda&1\\
0&0&\lambda
\end{pmatrix},
\eean
where the diagonal elements are filled with the eigenvalue $\lambda$ and the superdiagonal is composed of ones.

A left generalized eigenvector of rank $m$ corresponding to eigenvalue $\lambda$ satisfies
\begin{align}\label{eq:geneigvec}
    \vec v^{\text T}\left(\hat{\vec M}-\lambda \mathds{1}\right)^{m} & =0  \\
\intertext{and}
\label{eq:geneigvec2}
    \vec v^{\text T}\left(\hat{\vec M}-\lambda \mathds{1}\right)^{m-1} & \neq 0.
\end{align}
The parallel condition in Eq.~\eqref{eq:parallelcondition} is fulfilled by the left \textit{ordinary eigenvectors} denoted as $\vec b$ which are the rank-1 generalized left eigenvectors. The $i$th row in $\hat{\vec P}^{-1}$ is an ordinary left eigenvector of $\hat{\vec M}$ if the $i$th row in $\hat{\vec J}$ does not contain a superdiagonal 1 element. Linear combinations of ordinary eigenvectors corresponding to the same eigenvalue are also ordinary eigenvectors.

In the following subsections we provide the Jordan normal forms corresponding to the
Hamiltonians and 
degeneracy patterns discussed in subsections~\ref{sec:neutral}, and~\ref{sec:charged}, and section~\ref{sec:ellipsoid}.
Note that in these examples discussed in the main text, 
we set
both the interaction matrix $\hat{\vec R}=\mathds{1}_{3\times 3}$ and the right $g$-tensor $\hat{\vec g}_{\text R}=\mathds{1}_{3\times 3}$ as
the $3\times 3$ unit matrix, and
hence the matrix $\hat{\vec M}$ equals the left $g$-tensor $\hat{\vec g}_{\text L}$.

\subsection{Degeneracy pattern (II), 
section~\ref{sec:neutral}}

For the example Hamiltonian producing the
degeneracy pattern (II), 
treated in section~\ref{sec:neutral}, 
the left $g$-tensor was specified in
Eq.~\eqref{eq:neutralgtensors},
leading to
\bean\label{eq:jordan2a}
\hat{\vec M}_{\text{II}}=\hat{\vec g}_{\text {L,II}}\hat{\vec R}_{\text{II}}\hat{\vec g}_{\text {R,II}}^{-1}=\hat{\vec g}_{\text{L,II}}=\begin{pmatrix}
2&0&0\\
0&2&0\\
0&1&4
\end{pmatrix}.
\eean
The matrices of the Jordan decomposition 
[see Eq.~\eqref{eq:jordan}] of this matrix
$\hat{\vec M}_\text{II}$ read:
\bean\label{eq:jordan2b}
\hat{\vec J}_{\text{II}} =
\begin{pmatrix}
4&0&0\\
0&2&0\\
0&0&2
\end{pmatrix},\hspace{6mm}
\hat{\vec P}_{\text{II}} =
\frac{1}{2}
\begin{pmatrix}
0&0&2\\
0&2&0\\
1&-1&0
\end{pmatrix},\hspace{6mm}
\hat{\vec P}_{\text{II}}^{-1}=
\begin{pmatrix}
0&1&2\\
0&1&0\\
1&0&0
\end{pmatrix}.
\eean
The Jordan normal form $\hat{\vec J}_{\text{II}}$ 
is diagonal, it consists of 
three Jordan blocks: 
two blocks of the form 
$\hat{\vec J}_{1\times 1}(2)$,
and one block of the form 
$\hat{\vec J}_{1\times 1}(4)$.

As seen from Eqs.~\eqref{eq:jordan}
and \eqref{eq:jordan2b},
the left 
eigenvector corresponding to the eigenvalue 4 is $\vec v_1=(0,1,2)^T$. 
As claimed in Eq.~\eqref{eq:weylpoints},
there are two magnetic Weyl points,
forming a time-reversed pair, 
along the B-field direction set by $\vec v_1$.
It is also seen in Eq.~\eqref{eq:jordan2b}
that two left eigenvectors corresponding
to the eigenvalue 2 
are $\vec v_2 = (0,1,0)^T$
and $\vec v_3 = (1,0,0)^T$.
In fact, all vectors in 
the subspace $\operatorname{Span}(\vec v_2,\vec v_3)$
are eigenvectors with eigenvalue 2. 
According to Eqs.~\eqref{eq:bnorm}
and \eqref{eq:neutralgtensors},
the corresponding magnetic degeneracy points 
form a circle with radius $R=3/4$.

\subsection{Degeneracy pattern (V), section~\ref{sec:charged}}

In section~\ref{sec:charged},
describing the charged ellipse degeneracy pattern (V), 
the left $g$-tensor and hence
the matrix $\hat{\vec M}$ was set to
\bean\label{eq:jordan5a}
\hat{\vec M}_{\text{V}} =
\begin{pmatrix}
2&0&0\\
0&2&0\\
0&1&2
\end{pmatrix}.
\eean
A Jordan decomposition of this matrix is:
\bean\label{eq:jordan5b}
\hat{\vec J}_{\text{V}}=
\begin{pmatrix}
2&1&0\\
0&2&0\\
0&0&2
\end{pmatrix},\hspace{6mm}
\hat{\vec P}_{\text{V}}=
\begin{pmatrix}
0&0&1\\
0&1&0\\
1&0&0
\end{pmatrix},\hspace{6mm}
\hat{\vec P}_{\text{V}}^{-1}=
\begin{pmatrix}
0&0&1\\
0&1&0\\
1&0&0
\end{pmatrix}.
\eean
The Jordan normal form $\hat{\vec J}_{\text{V}}$ consists of the Jordan blocks $\hat{\vec J}_{1\times 1}(2)$ and $\hat{\vec J}_{2\times 2}(2)$. There is a superdiagonal 1 element in the first row, hence
$\vec v_1=(0,0,1)^\text{T}$ is not an ordinary left eigenvector
but a rank-2 generalized left eigenvector.

The ordinary eigenvectors are in the subspace $\operatorname{Span}(\vec v_2,\vec v_3)$ similarly to the previous case. 
As a consequence of this, and Eq.~\eqref{eq:bnorm}, 
the corresponding degeneracies are on a circle with radius $R=3/4$ in the $xy$-plane again.
This is similar to the degeneracy circle 
of pattern (II), but there the circle was
neutral, while here it is charged.

Using Figs.~\ref{fig:charged}~a,b,c, we argued in the main text
that the topological charge of this charged degeneracy circle
is localized on two opposite points of the circle. 
Here we claim that the charge is located in the $\vec v_2$ direction. 
This can be illustrated by studying a parameter-dependent
matrix $\hat{\vec M}$ that exemplifies a transition 
from the degeneracy pattern (II) formed by two equally-charged
Weyl points and a neutral circle, to the degeneracy pattern
(V) formed by the charged circle:
\bean
\hat{\vec{M}}_{\text{II}\rightarrow\text{V}}(\epsilon)=
\begin{pmatrix}
2&0&0\\
0&2&0\\
0&1&2+\epsilon
\end{pmatrix}.
\eean
For $\epsilon > 0$, this matrix has a Jordan decomposition
analogous to \eqref{eq:jordan2b}, i.e., it implies the
degeneracy pattern (II).
In particular, its largest eigenvalue is $2+\epsilon$,
with an ordinary left eigenvector $\vec v_1=(0,1,\epsilon)^{\text T}$.
As $\epsilon$ is tuned continuously to zero, then 
$\vec v_1$ coalesces with the 
($\epsilon$-independent)
second left ordinary eigenvector
$\vec v_2 = (0,1,0)^\text{T}$.
During this transition the Weyl points approach the neutral ellipse and then merge with it at $\epsilon=0$.

\subsection{Degeneracy pattern (IV), section~\ref{sec:ellipsoid}}

In section~\ref{sec:ellipsoid}, describing the charged
ellipsoid degeneracy pattern (IV), the
left $g$-tensor and hence the matrix $\hat{\vec M}$
was set to 
\bean\label{eq:jordan4}
\hat{\vec M}_{\text{IV}} =2\cdot\mathds{1}_{3\times 3}
\eean
Every vector is an eigenvector, so there are degeneracies in every direction. The distance of the degeneracy points from the origin is
given by Eq.~\eqref{eq:bnorm}

\section{Rank of the effective $g$-tensor in 
the points of the degeneracy ellipses}
\label{sec:effectivegtensor}

Figure~\ref{fig:neutral}
of the main text provides numerical evidence
that the topological charge density of the
neutral degeneracy ellipse, of pattern (II), 
is zero. 
Similarly, Fig.~\ref{fig:charged} shows that
the topological charge density of the 
charged degeneracy ellipse, pattern (V) 
is concentrated at two charged points, whereas
the linear charge density in all other points
of the ellipse is zero. 
In this section, we show that the above charge
distributions are related to the ranks of
the effective $g$-tensors $\hat{\vec g}_{\text{eff}}(\vec B_0)$ 
of the degeneracy points, 
defined in Eq.~\eqref{eq:Heff}.
Namely, the
effective $g$-tensor 
is a rank-1 matrix for the charged points of the charged ellipse, 
and a rank-2 matrix for the uncharged points of the charged ellipse
and for every point of the neutral ellipse, too. 
The rank-2 property implies a first-order 
energy splitting of the degeneracy as we leave the degenerate 
line in any perpendicular direction. 
In the case of a rank-1 effective $g$-tensor there exist 
a direction perpendicular to the 
plane of the degeneracy circle,
with the property that the energy splitting 
is of higher-than-linear order if we leave
the circle in that particular direction. 

The effective $g$-tensor for a ground-state 
degeneracy point
at $\vec B_0=B_0\vec b$ reads (Eq.~(E11) of Ref.~\onlinecite{Frank})
\bean\label{eq:geff1}
\hat{\vec g}_{\text{eff}}(\vec B_0)\hat{\vec O}=\left(\frac{\hat{\vec M}+a^2}{1+a^2}-\frac{\hat{\vec M}-a}{\sqrt{1+a^2}}\right)\hat{\vec g}_{\text{R}}\hat{\vec R}^{-1}\tilde{\vec b}\tilde{\vec b}^{\text T}+\frac{\hat{\vec M}-a}{\sqrt{1+a^2}}\hat{\vec g}_{\text{R}}\hat{\vec R}^{-1},
\eean
where
\bean
\label{eq:bbtilde}
\tilde{\vec b}=\hat{\vec R}\hat{\vec g}_{\text{R}}^{\text{T}}\vec b/|\hat{\vec g}_{\text{R}}^{\text{T}}\vec b|.
\eean
Note that, here, 
we use a slightly different notation,
compared to that in Ref.~\onlinecite{Frank}:
here we denote the dyadic product as a matrix product
$\tilde{\vec b}\tilde{\vec b}^{\text T}$ of
a column vector and a row vector, instead of
the alternative notation $\tilde{\vec b}\otimes\tilde{\vec b}$.
The orthogonal matrix $\hat{\vec O}$, 
which is defined [see Eq. (C2) of Ref.~\onlinecite{Frank}]
as an (ambiguous) rotation fulfilling $\hat{\vec O} \tilde{\vec b}
= \vec e_z$, e.g., the $\pi$ rotation 
around the bisector of $\tilde{\vec b}$ and $\vec e_z$.

Equation \eqref{eq:geff1} for the 
effective $g$-tensor is too complicated to determine 
$\operatorname{rank}\left[\hat{\vec g}_\text{eff}(\vec B_0)\right]$
directly.
Instead, in the forthcoming calculation, 
we show that the rank is reflected by the rank of a
a simpler matrix $\hat{\vec A}_Q$, 
which we express below in Eq.~\eqref{eq:Amatrix}. 
Then, in subsections~\ref{sec:neutraleffectiveg} 
and~\ref{sec:chargedeffectiveg},
we use this matrix $\hat{\vec A}_Q$ 
to derive the rank of the effective $g$-tensor on degeneracy ellipses.

As a first step in our calculation, 
we substitute $\tilde{\vec b}$ with $\vec b$
according to Eq.~\eqref{eq:bbtilde}, 
into Eq.~\eqref{eq:geff1}, and multiply 
the latter with $\hat{\vec O}^{-1}$ from the right: 
\bean\label{eq:geff2}
\hat{\vec g}_{\text{eff}}(\vec b)=\left[\left(\frac{\hat{\vec M}+a^2}{1+a^2}-\frac{\hat{\vec M}-a}{\sqrt{1+a^2}}\right)\frac{\hat{\vec g}_{\text{R}}\hat{\vec g}_{\text{R}}^{\text{T}}\vec b \vec b^{\text{T}}}{\vec b^{\text{T}}\hat{\vec g}_{\text{R}}\hat{\vec g}_{\text{R}}^{\text{T}}\vec b}+\frac{\hat{\vec M}-a}{\sqrt{1+a^2}}\right]\hat{\vec g}_{\text{R}}\hat{\vec R}^{-1}\hat{\vec O}^{-1}.
\eean
Here, we use $\vec b$ instead of $\vec B_0$ 
in the argument of $\hat{\vec g}_\text{eff}$, 
because the latter quantity does not depend on $B_0$.

In cases (II) and (V), where degeneracy ellipses appear,
the directions of the magnetic degeneracy points 
are in the subspace of the second and third left generalized 
eigenvectors.
That is, the degeneracy points
can be parameterized by the angle $\varphi \in [0,2\pi[$
via 
\bean\label{eq:bv2}
\vec b=\beta_2\vec v_2 +\beta_3\vec v_3=
\beta (\vec v_2 \cos\varphi + \vec v_3 \sin\varphi),
\eean
with $\beta=\sqrt{\beta_2^2+\beta_3^2}$. 
Since $\vec v_2$ and $\vec v_3$ are not necessarily orthogonal and normalized, $\beta$ is not necessarily $|\vec b|$ and $\varphi$ is not necessarily the angle between $\vec b$ and $\vec v_2$. 

The forthcoming steps lack an a priori intuitive justification, 
but a posteriori they prove to be particularly useful. 
First, let us recall that according to Eq.~\eqref{eq:jordan}, 
the left generalized eigenvectors 
form the rows in the similarity 
transformation matrix
$\hat{\vec{P}}^{-1}$
in the Jordan decomposition of $\hat{\vec M}$. 
Using this, $\vec b^{\text T}$ can be written as the second row of the matrix $\hat{\vec Q}\hat{\vec P}^{-1}$, 
that is
\bean
\label{eq:bT}
\vec{b}^\text{T} = \left(
    \hat{\vec Q} \hat{\vec P}^{-1}
\right)_{2,.},
\eean
where
\bean\label{eq:Q}
\hat{\vec Q}=\beta
\begin{pmatrix}
1&0&0\\
0&\cos\varphi&\sin\varphi\\
0&-\sin\varphi&\cos\varphi
\end{pmatrix}.
\eean

Inserting a unit matrix in the form of 
$\hat{\vec P}^{-1} \hat{\vec P}$ to 
Eq.~\eqref{eq:bT} yields
\bean
\label{eq:bv2Q}
\vec b^\text{T} 
= 
\left(\hat{\vec P}^{-1}
\hat{\vec P}
\hat{\vec Q} \hat{\vec P}^{-1}\right)_{2,.}
\equiv 
\left(\hat{\vec P}^{-1} \hat{\vec Q}'\right)_{2,.}
= \vec{v}_2^{\text{T}} \hat{\vec Q}'.
\eean
Here the sign $\equiv$ denotes the definition of $\hat{\vec Q}'$
and in the last step, we used
Eq.~\eqref{eq:jordan}.

Substituting Eq.~\eqref{eq:bv2Q} to 
the effective $g$-tensor of Eq.~\eqref{eq:geff2},
we obtain
\bean\label{eq:geff3}
\hat{\vec g}_{\text{eff}}(\hat{\vec Q}'^{\text T}\vec v_2)=\left[\left(\frac{\hat{\vec M}+a^2}{1+a^2}-\frac{\hat{\vec M}-a}{\sqrt{1+a^2}}\right)
\hat{\vec D}
+\frac{\hat{\vec M}-a}{\sqrt{1+a^2}}\right]\hat{\vec g}_{\text{R}}\hat{\vec R}^{-1}\hat{\vec O}^{-1},
\eean
where we introduced the shorthand
\bean\label{eq:d}
\hat{\vec D}=\frac{\hat{\vec g}_{\text{R}}\hat{\vec g}_{\text{R}}^{\text{T}}\hat{\vec Q}'^{\text T}\vec v_2 \vec v_2^{\text{T}}\hat{\vec Q}'}{\vec v_2^{\text{T}}\hat{\vec Q}'\hat{\vec g}_{\text{R}}\hat{\vec g}_{\text{R}}^{\text{T}}\hat{\vec Q}'^{\text T}\vec v_2},
\eean
which is a dyadic product. 

Next, we further transform $\hat{\vec M}$ and
$\hat{\vec D}$ in Eq.~\eqref{eq:geff3}, starting with
the latter. 
Substituting unit matrices into Eq.~\eqref{eq:d}
yields
\bean\label{eq:d2}
\hat{\vec D}=\frac{\left(\hat{\vec Q}'^{-1}\hat{\vec P}\hat{\vec P}^{-1}\hat{\vec Q}'\right)\hat{\vec g}_{\text{R}}\hat{\vec g}_{\text{R}}^{\text{T}}\hat{\vec Q}'^{\text T}\vec v_2 \vec v_2^{\text{T}}\hat{\vec Q}'}{\vec v_2^{\text{T}}\left(\hat{\vec P}\hat{\vec P}^{-1}\right)\hat{\vec Q}'\hat{\vec g}_{\text{R}}\hat{\vec g}_{\text{R}}^{\text{T}}\hat{\vec Q}'^{\text T}\vec v_2}.
\eean
Then, using the associative nature
of matrix multiplication, we obtain 
\bean\label{eq:d3}
\hat{\vec D}=\frac{\hat{\vec Q}'^{-1}\hat{\vec P}\left(\hat{\vec P}^{-1}\hat{\vec Q}'\hat{\vec g}_{\text{R}}\hat{\vec g}_{\text{R}}^{\text{T}}\hat{\vec Q}'^{\text T}\vec v_2\right) \vec v_2^{\text{T}}\hat{\vec Q}'}{\vec v_2^{\text{T}}\hat{\vec P}\left(\hat{\vec P}^{-1}\hat{\vec Q}'\hat{\vec g}_{\text{R}}\hat{\vec g}_{\text{R}}^{\text{T}}\hat{\vec Q}'^{\text T}\vec v_2\right)}=\frac{\hat{\vec Q}'^{-1}\hat{\vec P}\vec r \vec v_2^{\text{T}}\hat{\vec Q}'}{\vec v_2^{\text{T}}\hat{\vec P}\vec r},
\eean
where we introduced
\bean
\label{eq:rvector}
\vec r=\hat{\vec P}^{-1}\hat{\vec Q}'\hat{\vec g}_{\text{R}}\hat{\vec g}_{\text{R}}^{\text{T}}\hat{\vec Q}'^{\text T}\vec v_2.
\eean
Then, we substitute the definition of $\hat{\vec Q}'$ from Eq.~\eqref{eq:bv2Q} to Eq.~\eqref{eq:d3}, 
yielding 
\bean\label{eq:d4}
\hat{\vec D}=\frac{\hat{\vec P}\hat{\vec Q}^{-1}\hat{\vec P}^{-1}\hat{\vec P}\vec r \vec v_2^{\text{T}}\hat{\vec P}\hat{\vec Q}\hat{\vec P}^{-1}}{\vec v_2^{\text{T}}\hat{\vec P}\vec r}=\hat{\vec P}\hat{\vec Q}^{-1}\frac{\vec r}{r_2}
\begin{pmatrix}
0&1&0
\end{pmatrix}
\hat{\vec Q}\hat{\vec P}^{-1}.
\eean
The denominator $r_2$, which is the second
vector component of $\vec r$, 
appears from the scalar product of \mbox{$\vec r$ with $\vec v^{\text T}_2\hat{\vec P}\equiv(0\ 1\ 0)$}. 
Note that $r_2 = |\hat{\vec g}_{\text R}^{\text T}\vec b|^2$,
which follows, e.g., from Eqs.~\eqref{eq:rvector} and 
\eqref{eq:bv2Q}, 
and it guarantees that the denominator in 
Eq.~\eqref{eq:d4} is nonzero.

According to Eq.~\eqref{eq:d4}, 
the matrix $\hat{\vec D}$ can be thought of 
as a result of a similarity
transformation generated by 
$\hat{\vec P}\hat{\vec Q}^{-1}$.
Now, we transform the terms 
of Eq.~\eqref{eq:geff3}
containing $\hat{\vec M}$ to 
a similar form.
Multiplying $\hat{\vec M}$ with appropriately
composed unit matrices, using its Jordan
decomposition, 
and introducing the \emph{transformed Jordan normal form}
via
\bean\label{eq:JQ}
\hat{\vec J}_Q=\hat{\vec Q}\hat{\vec J}\hat{\vec Q}^{-1},
\eean
we find
\bean
\label{eq:mresult}
\hat{\vec M}=\left(\hat{\vec P}\hat{\vec Q}^{-1}\hat{\vec Q}\hat{\vec P}^{-1}\right)\left(\hat{\vec P}\hat{\vec J}\hat{\vec P}^{-1}\right)\left(\hat{\vec P}\hat{\vec Q}^{-1}\hat{\vec Q}\hat{\vec P}^{-1}\right)=\hat{\vec P}\hat{\vec Q}^{-1}
\hat{\vec J}_Q
\hat{\vec Q}\hat{\vec P}^{-1}.
\eean

Inserting Eqs.~\eqref{eq:d4} and \eqref{eq:mresult} 
into Eq.~\eqref{eq:geff3}, we find
the following expression for the
effective $g$-tensor: 
\bean\label{eq:geff5}
\hat{\vec g}_{\text{eff}}(\vec b)=\hat{\vec P}\hat{\vec Q}^{-1}
\hat{\vec{A}}_Q
\hat{\vec Q}\hat{\vec P}^{-1}\hat{\vec g}_{\text{R}}\hat{\vec R}^{-1}\hat{\vec O}^{-1},
\eean
where we introduced
\bean\label{eq:Amatrix}
\hat{\vec A}_{Q}=\left(\frac{\hat{\vec J}_Q+a^2}{1+a^2}-\frac{\hat{\vec J}_Q-a}{\sqrt{1+a^2}}\right)\frac{\vec r}{r_2}
\begin{pmatrix}
0&1&0
\end{pmatrix}+\frac{\hat{\vec J}_Q-a}{\sqrt{1+a^2}}.
\eean
Matrix $\hat{\vec A}_Q$ has the same rank as the effective $g$-tensor, because they only differ by multiplications of non-singular matrices.
In what follows, we will determine 
the rank of the $g$-tensor at the points
of the degeneracy ellipses by determining the 
rank of $\hat{\vec A}_Q$.

\subsection{Neutral ellipse}
\label{sec:neutraleffectiveg}

In Eq.~\eqref{eq:jordan2b}, 
we have shown an example Jordan decomposition
corresponding to a degeneracy pattern 
including a neutral ellipse. 
More generally, the normal form 
of that degeneracy pattern has
a two-fold degeneracy of the following kind
[see Table I of Ref.~\onlinecite{Frank}]:
\bean\label{eq:jordanneutral}
\hat{\vec J}_{\text{II}}=\begin{pmatrix}
b&0&0\\
0&a&0\\
0&0&a
\end{pmatrix},
\eean
where $a,b>0$ and $a\neq b$. 
Since its second Jordan block is proportional to the $2\times 2$ unit matrix, the transformation with $\hat{\vec Q}$ defined in Eq.~\eqref{eq:Q} leaves the normal form invariant:
\bean\label{eq:JQneutral}
\hat{\vec J}_{Q\text{,II}}=\hat{\vec J}_{\text{II}},
\eean
for every $\varphi$. That means that every point of a neutral ellipse has the same rank effective $g$-tensor. 
Then, expressing $\hat{\vec A}_Q$ 
from Eqs.~\eqref{eq:Amatrix}, \eqref{eq:JQneutral}
and \eqref{eq:jordanneutral} yields
\begin{equation}\label{eq:Amatrix2}
\begin{split}
\hat{\vec A}_Q&=
\begin{pmatrix}
\frac{b+a^2}{1+a^2}-\frac{b-a}{\sqrt{1+a^2}}&0&0\\
0&\frac{a+a^2}{1+a^2}&0\\
0&0&\frac{a+a^2}{1+a^2}
\end{pmatrix}
\begin{pmatrix}
r_1/r_2\\1\\r_3/r_2
\end{pmatrix}
\begin{pmatrix}
0&1&0
\end{pmatrix}+
\begin{pmatrix}
\frac{b-a}{\sqrt{1+a^2}}&0&0\\
0&0&0\\
0&0&0
\end{pmatrix}
\\
&=\begin{pmatrix}
\frac{b-a}{\sqrt{1+a^2}}&\left[\frac{b+a^2}{1+a^2}-\frac{b-a}{\sqrt{1+a^2}}\right]\frac{r_1}{r_2}&0\\
0&\frac{a(1+a)}{1+a^2}&0\\
0&\frac{a(1+a)}{1+a^2}\frac{r_3}{r_2}&0
\end{pmatrix}.
\end{split}
\end{equation}
$\hat{\vec A}_{Q}$ cannot be a dyadic product because the condition $A_{Q,11}A_{Q,22}=A_{Q,12}A_{Q,21}$ cannot be satisfied as  \mbox{$A_{Q,11},A_{Q,22}\neq 0$} and $A_{Q,21}=0$. That means $\operatorname{rank}(\hat{\vec g}_\text{eff})>1$. The determinant 
of the effective $g$-tensor 
is zero [Eq.~(E23) of Ref.~\onlinecite{Frank} with $c=a$],
therefore $\operatorname{rank}(\hat{\vec g}_\text{eff})<3$. 
This way we proved that the rank is 2 for every points of the neutral degeneracy ellipse.

\subsection{Charged ellipse}
\label{sec:chargedeffectiveg}

In Eq.~\eqref{eq:jordan5b}, we have shown
an example for a Jordan decomposition corresponding to a 
degeneracy pattern of a charged ellipse. 
More generally, the Jordan normal form of that
degeneracy pattern has a three-fold eigenvalue degeneracy,
and a single 1 element in the superdiagonal:
\bean\label{eq:chargedjordan}
\hat{\vec J}_{\text{V}}=
\begin{pmatrix}
a&1&0\\
0&a&0\\
0&0&a
\end{pmatrix},
\eean
with $a>0$. 
Now $\hat{\vec J}_{Q\text{,V}}$ does depend
on the angle $\varphi$ parameterizing the degeneracy
point along the degeneracy ellipse:
\bean\label{eq:JQcharged}
\hat{\vec J}_{Q\text{,V}} =
\begin{pmatrix}
a&\cos\varphi&-\sin\varphi\\
0&a&0\\
0&0&a
\end{pmatrix}.
\eean
Expressing $\hat{\vec A}_Q$ 
from Eqs.~\eqref{eq:Amatrix} and \eqref{eq:JQcharged}
yields
\begin{equation}
\begin{split}
\label{eq:Amatrix3}
\hat{\vec A}_Q&=
\begin{pmatrix}
\frac{a+a^2}{1+a^2}&\cos\varphi\left[\frac{1}{1+a^2}-\frac{1}{\sqrt{1+a^2}}\right]&-\sin\varphi\left[\frac{1}{1+a^2}-\frac{1}{\sqrt{1+a^2}}\right]\\
0&\frac{a+a^2}{1+a^2}&0\\
0&0&\frac{a+a^2}{1+a^2}
\end{pmatrix}
\begin{pmatrix}
r_1/r_2\\1\\r_3/r_2
\end{pmatrix}
\begin{pmatrix}
0&1&0
\end{pmatrix}+
\begin{pmatrix}
0&\frac{\cos\varphi}{\sqrt{1+a^2}}&\frac{-\sin\varphi}{\sqrt{1+a^2}}\\
0&0&0\\
0&0&0
\end{pmatrix}
\\
&=\begin{pmatrix}
0&A_{Q,12}&-\frac{\sin(\varphi)}{\sqrt{1+a^2}}\\
0&\frac{a(1+a)}{1+a^2}&0\\
0&A_{Q,32}&0
\end{pmatrix},
\end{split}
\end{equation}
where the elements $A_{Q,12}$ and $A_{Q,32}$ 
are given by lengthy but unimportant expressions.
Similarly to the neutral ellipse, the condition  $A_{Q,12}A_{Q,23}=A_{Q,13}A_{Q,22}$  is not satisfied for $\sin\varphi\neq0$, hence 
the rank of the effective $g$-tensor 
is 2 for those points. 
However, if $\sin\varphi=0$, i.e., if the magnetic
field is along the direction of $\vec v_2$, the matrix $\hat{\vec A}$ is clearly a dyadic product and not a zero matrix,
hence its rank is 1, implying that the rank 
of the effective $g$-tensor is also 1.

Finally, let us consider this latter case, 
when the rank of the effective $g$-tensor is 1. 
Starting at the degeneracy point $\vec B_0$,
and changing the magnetic field by $\delta \vec B$
along the degeneracy line as
$\vec B = \vec B_0 + \delta \vec B$,
the energy splitting induced by $\delta \vec B$ 
is at least of second order in $\delta \vec B$.
Since the rank of the effective $g$-tensor is 1, 
there must be a plane of higher-order splitting, 
that is, a plane along which 
the energy splitting is at least of second order
in $\delta \vec B$. 
Which is the second direction, which spans
this plane together with the direction of the 
degeneracy ellipse? 

This question can be answered by recasting
the rank-1 effective $g$-tensor as a dyadic product
of two vectors.
Without the derivation, we claim that one way
this can be done is as follows:
\bean\label{eq:dyadicgeff}
\hat{\vec g}_{\text{eff}}(\vec v_2)=\vec d_1\vec d_2^{\text{T}},
\eean
where
\bean\label{eq:dyadicgeff2}
\vec d_1=\frac{1}{1+a^{2}}\left[a(1+a)\frac{\hat{\vec g}_{\text{R}}\hat{\vec g}_{\text{R}}^{\text T}\vec v_{2}}{\vec v_{2}^{\text{T}}\hat{\vec g}_{\text{R}}\hat{\vec g}_{\text{R}}^{\text T}\vec v_{2}}+\vec w_{1}
\right],\hspace{6mm}
\vec d_2=\hat{\vec O}\hat{\vec R}\hat{\vec g}_{\text{R}}^{\text T}\vec v_2
\eean
with $\vec w_1$ is the right generalized eigenvector defined in Eq.~\eqref{eq:jordan}. 
The column vector $\vec d_1$ in the square bracket defines the direction 
of maximal linear splitting in $\delta \vec B$, cf. Eq.~\eqref{eq:Heff}.
If $\delta \vec B$ lies in the plane perpendicular to 
that vector, then the energy splitting is
at least second order in $\delta \vec B$.

\section{Rank-2 points of a degeneracy line carry zero linear topological
charge density}
\label{sec:zerodensity}

In this section, we show in general that degeneracy lines consisting of rank-2 points carry zero linear topological charge density.
For simplicity, we assume that the degeneracy line is along the $B_z$ axis.

We use $x\equiv B_x$, 
$y \equiv B_y$ and $z \equiv B_z$ for brevity, 
and we shift the coordinate system of the magnetic parameter
space such that the rank-2 point we consider is in the 
origin, where $(x,y,z) =0$.
From now on, we further simplify notation
by using $\psi_0(s,\vartheta)$ instead
of $\tilde{\psi}_0(s,\vartheta)$
(cf. Eq.~\eqref{eq:2dalternative}).

According to Eq.~\eqref{eq:apparent}, 
our goal is to evaluate 
\begin{equation}
\label{eq:rank2zerocharge}
    \nu(0) = \frac{1}{2\pi} \lim_{r \to 0}
    \int_0^{2\pi} d\vartheta
    \mathcal{B}_\text{2D}(s,\vartheta)
    =
    \frac{1}{2\pi}
    \lim_{r \to 0}
    \int_0^{2\pi} d\vartheta
    (-2) \operatorname{Im}\left[
        \braket{\partial_\vartheta \psi_0 | \partial_s \psi_0}
    \right].
\end{equation}
For the parameter-space geometry we consider, 
the relation between the Cartesian coordinates
and the torus parameters is 
$x = r \cos\vartheta$, $y = r \sin \vartheta$, and $s = z$.
Our strategy is to evaluate the above $\vartheta$ integral, 
i.e., to show that it vanishes, by 
using an approximate ground state $\psi_0$ 
for $r \to 0$, obtained via a $z$-dependent
two-level effective Hamiltonian. 

Along the $z$ axis, in the small neighborhood
of the degeneracy point, we take an orthonormal basis
$(\eta(z),\chi(z))$
of the degenerate ground-state subspace for each $z$,
such that the two basis states depend on $z$ continuously. 
Using this basis, we define the effective Hamiltonian 
of the degeneracy point $(0,0,z)$ as follows:
\begin{equation}
    H_\text{eff}(z) = P(z) H(x,y,z) P(z),
\end{equation}
where $P(z) = \ket{\eta(z)}\bra{\eta(z)} + \ket{\chi(z)} \bra{\chi(z)}$
projects on the two-dimensional ground-state subspace.
(This projector $P$ should not be confused by 
the similarity transformation $\hat{\vec P}$.)
This effective Hamiltonian can also be written as
\begin{equation}
    H_\text{eff}(z) = P(z) (H_0 + z H_z) P(z) 
    + P(z) (x H_x + y H_y) P(z), 
\end{equation}
since our Hamiltonian [Eq.~\eqref{eq:hamiltonian}] 
is a linear function of the 
magnetic field coordinates $x$, $y$, $z$.
Furthermore, by definition, $P(z)$ projects to the twofold degenerate
ground-state subspace of $H_0 + z H_z$. 
Therefore, if we drop the degenerate part of the effective
Hamiltonian, then we obtain 
\begin{equation}
    H_\text{eff}(z) = 
    x P(z) H_x P(z) + y P(z) H_y P(z).
\end{equation}

Up to now, the basis $(\eta(z), \chi(z))$ was
ambigous, its defining constraints being that it has to 
depend continuously on $z$, and it has to 
span the two-dimensional ground-state subspace of $H_0 + z H_z$.
Now, we further restrict this basis such that 
$\braket{\eta(z)| H_x | \eta(z)} =
\braket{\chi(z)| H_x | \chi(z)} = 0$,
$\braket{\eta(z)| H_y | \eta(z)} =
\braket{\chi(z)| H_y | \chi(z)} = 0$,
and $g_{xx}(z) = \braket{\eta(z)| H_x | \chi(z)} > 0$.
Then, the effective Hamiltonian is written as
\begin{equation}
    H_\text{eff}(z) = x g_{xx}(z) \tau_x(z)+
    y g_{yx}(z) \tau_x(z) + y g_{yy}(z) \tau_y(z),
\end{equation}
with $g_{yx}(z) = \text{Re} \left[\braket{\eta(z)| H_y | \chi(z)} \right]$
and $g_{yy}(z) = -\operatorname{Im} \left[\braket{\eta(z)| H_y | \chi(z)} \right]$.
Note that the rank-2 character of the origin guarantees
that $g_{yy}(z) \neq 0$.

This two-level effective Hamiltonian
is straightforward to diagonalize, and its diagonalization
provides
a formula for the unique ground state away from the $z$ axis:
\begin{equation}
\ket{\psi_0(s,\vartheta) }
\equiv
\ket{\psi_0(z,\vartheta) }
\approx 
\frac{1}{\sqrt{2}}
\left(
     \ket{\eta (z)} 
    -
    e^{i\alpha(z,\vartheta)} \ket{\chi(z)}
\right),
\label{eq:groundstate}
\end{equation}
where $\alpha(z,\vartheta)$ is the angle enclosed by 
the vectors $(1,0)$ and
$(g_{xx}(z) \cos \vartheta + g_{yx}(z) \sin \vartheta,
g_{yy}(z) \cos \vartheta)$.
A key property of this angle is
\begin{equation}
\label{eq:alpha1}
\alpha(z,\vartheta+\pi) = \alpha(z,\vartheta) + \pi,
\end{equation}
implying 
\begin{equation}
e^{i\alpha(z,\vartheta+\pi)} 
=
- e^{i\alpha(z,\vartheta)}.
\label{eq:alpha2}
\end{equation}
As we show below, this property implies
$\mathcal{B}_\text{2D}(s,\vartheta) = -
\mathcal{B}_\text{2D}(s,\vartheta + \pi)$,
and hence a vanishing result of the integral
in Eq.~\eqref{eq:rank2zerocharge}.

To compute the integrand of
Eq.~\eqref{eq:rank2zerocharge}, we first
evaluate the derivatives of the ground
state \eqref{eq:groundstate}:
\bean
\ket{\partial_\vartheta \psi_0(z,\vartheta)}&=&\frac{-i}{\sqrt{2}}(\partial_\vartheta \alpha(z,\vartheta)) e^{i\alpha(z,\vartheta)}\ket{\chi(z)},
\\
\ket{\partial_z \psi_0(z,\vartheta)} 
&=&\frac{1}{\sqrt{2}}\left(
\ket{\partial_z \eta(z,\vartheta)}
-
e^{i\alpha(z,\vartheta)}
\ket{\partial_z \chi(z)}
-i
(\partial_z \alpha(z,\vartheta)) e^{i\alpha(z,\vartheta)}\ket{\chi(z)}\right).
\eean
Then, the scalar product in the integrand
of Eq.~\eqref{eq:rank2zerocharge} reads
\bean
\braket{\partial_\vartheta \psi_0 | \partial_z \psi_0}=
\frac{1}{2} \left[
i(\partial_\vartheta \alpha(z,\vartheta)
)e^{-i\alpha(z,\vartheta)}
\braket{\chi(z)|\partial_z\eta(z)}
+i(\partial_\vartheta \alpha(z,\vartheta))\braket{\chi(z)|\partial_z\chi(z)}
+
(\partial_\vartheta \alpha(z,\vartheta))
(\partial_z \alpha(z,\vartheta))
\right].
\eean
The second term in the square bracket
is real, since
\bean
\text{Re} \braket{\chi(z)|\partial_z\chi(z)}
= \frac{1}{2} \left(
\braket{\chi(z)|\partial_z\chi(z)}
+
\braket{\partial_z \chi(z)|\chi(z)}
\right) = 
\frac{1}{2} \partial_z 
\braket{\chi(z) | \chi(z) } = 0.
\eean
Furthermore, the third term is 
also real, since $\alpha$, defined as
an angle, is real-valued. 

As a consequence, the imaginary 
part of the scalar product in the integrand
of Eq.~\eqref{eq:rank2zerocharge}
reads
\bean
\operatorname{Im} \braket{\partial_\vartheta \psi_0(z,\vartheta) | \partial_z \psi_0(z,\vartheta)}=
\frac{1}{4}
\left[
(\partial_\vartheta \alpha(z,\vartheta)
)e^{-i\alpha(z,\vartheta)}
\braket{\chi(z)|\partial_z\eta(z)}
+ c.c.
\right].
\eean
From this result, it follows that
\begin{equation}
\label{eq:alpha3}
\operatorname{Im} \braket{\partial_\vartheta \psi_0(z,\vartheta+\pi) | \partial_z \psi_0(z,\vartheta+\pi)}
    =
-  
\operatorname{Im} \braket{\partial_\vartheta \psi_0(z,\vartheta) | \partial_z \psi_0(z,\vartheta)}
\end{equation}
where we used Eqs.~\eqref{eq:alpha1} and
\eqref{eq:alpha2}.
Finally, Eq.~\eqref{eq:alpha3} implies that
the $\vartheta$ integral of
Eq.~\eqref{eq:rank2zerocharge}
vanishes. 

While the topological charge density vanishes at rank-2 points of a degeneracy line, the Berry curvature, even at an isolated degeneracy line, is not identically zero.
To see this, we consider a small disk $D$ intersecting the degeneracy line and calculate the integral of the Berry connection $\vec{\mathcal{A}} = i \braket{\psi_0 \vert \vec{\nabla} \psi_0}$ around the perimeter of the disk, which is a loop surrounding the line degeneracy:
\begin{equation}
    \oint_{\partial D} \vec{\mathcal{A}} \cdot d\vec{l} = i \int_0^{2\pi} \braket{\psi_0 | \partial_{\vartheta} \psi_0} d\vartheta = \frac{1}{2} \int_0^{2\pi} \partial_{\vartheta} \alpha(z, \vartheta) d\vartheta = \frac{1}{2} \left[\alpha(z, 2\pi) - \alpha(z, 0)\right] = \pi.
\end{equation}
This result is independent of the radius of the disk chosen.

If the degeneracy is broken by some perturbation, the Berry curvature becomes well defined and finite everywhere, with a large Berry curvature along the position of the degeneracy line.
In this case the integral of the Berry curvature for the disk equals the integral of the Berry connection around the perimeter
\begin{equation}
    \iint_D \vec{\mathcal{B}}\cdot d\vec{A} = \oint_{\partial D} \vec{\mathcal{A}} \cdot d\vec{l}.
\end{equation}
The connection integral changes continuously with perturbations of the Hamiltonian, while the curvature diverges at the degeneracy.
This allows us to interpret this result in the degenerate case as half a quantum of Berry flux concentrated in a linelike flux tube along the line degeneracy.

\section{Example for finite linear topological charge density}\label{sec:finitelinear}

In Eq.~\eqref{eq:topologicalchargedensity} we defined the linear topological charge density but showed later that it is either zero,
or is concentrated to single points akin to a Dirac delta. 
Here, we provide an example Hamiltonian with a degeneracy
line in its parameter space, such that the 
degeneracy line carries a finite, continuously-varying 
linear topological charge density.

Our example is a spin-1/2 Hamiltonian which is a nonlinear
function of its parameters $\vec B = (B_x,B_y,B_z)$:
\bean\label{eq:finitelinearH}
H(\vec B)&=&\left(B_x^2-B_y^2\right)S_x+2B_xB_yS_y+\left(B_x^2+B_y^2\right)B_z S_z=\vec B_{\text{eff}}\cdot\boldsymbol{S}.
\eean
We will call the quantity $\vec B_\text{eff}$ the \emph{effective magnetic field}.
The Hamiltonian in Eq.~\eqref{eq:finitelinearH} 
has a degeneracy line along the $B_z$ axis. 
For small but finite $B_x$ and/or $B_y$,
the degenerate ground state splits in energy, 
quadratically in $B_x$ and $B_y$.
To calculate the linear topological charge density of
the degeneracy line along the $B_z$ axis, 
we follow the route introduced in section~\ref{sec:linear}, 
utilizing Eq.~\eqref{eq:topologicalchargedensity}.

We consider a cylinder of finite radius $r$ surrounding 
the $B_z$ axis. 
We parametrize the points of this cylinder with cylindrical coordinates,
via $(B_x,B_y,B_z) = (r \cos \vartheta, r \sin\vartheta, s)$.
At a given point of this cylinder, specified by $(r,\vartheta,s)$, 
the effective magnetic field reads:
\bean\label{eq:finitelinearBeff}
\vec B_{\text{eff}}=\begin{pmatrix}
r^2\cos^2\vartheta-r^2\sin^2\vartheta\\
2r^2\cos\vartheta \sin\vartheta\\
(r^2\cos^2\vartheta+r^2\sin^2\vartheta)s
\end{pmatrix}=\begin{pmatrix}
r^2\cos 2\vartheta\\
r^2\sin 2\vartheta\\
r^2 s
\end{pmatrix}.
\eean
The ground state can be expressed as
\bean\label{eq:finitelinearGS}
\ket{\psi_0}=\begin{pmatrix}
\sin\frac{\vartheta_{\text{eff}}}{2}\\
-e^{i\varphi_{\text{eff}}}\cos\frac{\vartheta_{\text{eff}}}{2}
\end{pmatrix},
\eean
where
\bean\label{eq:finitelinearpolar2}
\varphi_{\text{eff}}&=&2\vartheta, \\
\vartheta_{\text{eff}}&=&\tan^{-1} \left(\frac{1}{s}\right)
\eean
are the spherical angles of the effective magnetic field. 
For this specific Hamiltonian, these angles do not depend on the radius $r$,
hence the $r\to 0$ limit of Eq.~\eqref{eq:topologicalchargedensity} will
be omitted below. 

Having the $\vartheta$- and $s$-dependence of the ground
state $\ket{\psi_0}$ at hand, it is straightforward to calculate
the two-dimensional Berry curvature according to 
Eq.~\eqref{eq:2dberrycurvature}:
\bean\label{eq:finitelinearberry}
\mathcal{B}_{\text{2D}}(s,\vartheta)=-2\operatorname{Im}\braket{\partial_\vartheta\psi|\partial_s\psi}
=\frac{1}{\sqrt{(s^2+1)^3}},
\eean
which depends only on $s$. 
From this, using Eq.~\eqref{eq:topologicalchargedensity}, 
we evaluate the linear topological charge density:
\bean
\nu(B_z) \equiv \nu(s) 
=\frac{1}{\sqrt{(s^2+1)^3}},
\eean
which is indeed finite and depends continuously on the coordinate
$B_z \equiv s$ along the degeneracy line.

\section{Surface charge density of the charged ellipsoid}
\label{sec:surfacederivation}
Here, we derive the surface charge density of the
charged ellipsoid, a result quoted in the main 
text as Eq.~\eqref{eq:surfacechargedensity}. To do this, first, we transform the Hamiltonian to a simple form, where the Berry curvature is easy to determine, then we transform it back to obtain the surface charge density.
\subsection{Berry curvature in a simplified Hamiltonian}
The Hamiltonian introduced in Eq.~\eqref{eq:hamiltonian} can be simplified with the following steps. A global unitary transformation $U$ which changes the right spin as $\vec S_{\text R}'=\hat{\vec R}\vec S_{\text R}$ changes the interaction to be isotropic
\bean\label{eq:Hisoint}
H'(\vec B)=UH(\vec B)U^\dagger=\vec B\cdot\left(\hat{\vec g}_{\text L}\vec S_{\text L}+\hat{\vec g}_{\text R}\hat{\vec R}^{-1}\vec S_{\text R}\right)+J\vec S_{\text L}\cdot\vec S'_{\text R}.
\eean
This transformation changes the right $g$-tensor too. Now, we simplify the Zeeman term of the right spin with a linear transformation
\bean\label{eq:Btransform}
\vec B'=\hat{\vec R}\hat{\vec g}_{\text R}^{\text T}\vec B.
\eean
To do this, we substitute the unit matrix $\hat{\vec g}_{\text R}\hat{\vec R}^{-1}\hat{\vec R}\hat{\vec g}_{\text R}^{-1}$, we get
\bean\label{eq:HBtransform}
H'(\vec B)=\vec B\cdot\hat{\vec g}_{\text R}\hat{\vec R}^{-1}\left(\hat{\vec R}\hat{\vec g}_{\text R}^{-1}\hat{\vec g}_{\text L}\vec S_{\text L}+\vec S_{\text R}\right)+J\vec S_{\text L}\cdot\vec  S'_{\text R}.
\eean
This changes the left $g$-tensor to a transformed $\hat{\vec M}$ matrix
\bean\label{eq:Mtransform}
\hat{\vec R}\hat{\vec g}_{\text R}^{-1}\hat{\vec g}_{\text L}=\hat{\vec g}_{\text L}^{-1}\hat{\vec g}_{\text L}\hat{\vec R}\hat{\vec g}_{\text R}^{-1}\hat{\vec g}_{\text L}=\hat{\vec g}_{\text L}^{-1}\hat{\vec M}\hat{\vec g}_{\text L}=\hat{\vec M}'.
\eean
The result is a Hamiltonian with $\hat{\vec R}'=\mathds{1}_{3\times 3}$ and $\hat{\vec g}'_{\text R}=\mathds{1}_{3\times 3}$
\bean\label{eq:Hsimple}
H'(\vec B')=\vec B'\cdot\left(\hat{\vec M}'\vec S_{\text L}+\vec S_{\text R}\right)+J\vec S_{\text L}\cdot \vec S'_{\text R}.
\eean
The global unitary transformation leaves the Berry curvature invariant but the linear transformation in the parameter space changes it, as we derive it in the next subsection.
 
For degeneracy ellipsoids $\hat{\vec M} = a \mathds{1}_{3\times 3}$ is proportional to the unit matrix,
\bean\label{eq:Hsphere}
H'(\vec B')=\vec B'\cdot\left(a\vec S_{\text L}+\vec S_{\text R}\right)+J\vec S_{\text L}\cdot \vec S'_{\text R}.
\eean
The result is an isotropic Hamiltonian. It has a degeneracy sphere with radius
\bean\label{eq:Rsphere}
R=\frac{1}{2}\left(1+\frac{1}{a}\right).
\eean
carrying a total topological charge 2. The Berry curvature in the transformed parameter space can be calculated using Gauss's law because of the isotropy
\bean\label{eq:isoBerry}
  \mathbfcal{B}'(\vec B') =
    \begin{cases}
      0 & B'<R\\
      \frac{\vec B'}{B'^3} & B'>R.
    \end{cases}
\eean
\subsection{Transformation of the Berry curvature in 3 dimensional parameter space}
To obtain an isotropic Hamiltonian we did a global unitary transformation on the Hilbert space which preserves the Berry curvature, but we also did a linear transformation on the parameter space which, however, changes the curvature. In this subsection we derive the transformation of the Berry curvature in 3-dimensional parameter spaces.

We assume that Berry curvature with the transformed argument $\vec x'(\vec x)$ is known, this transformed curvature is given by $\mathcal{B}'_i\left(\vec x'\right) \equiv i\epsilon_{ijk} \braket{\partial'_{j}\psi|\partial'_{k}\psi}$ where $\partial'_k$ is the derivative with respect to $x'_k$.
We want to find the curvature with respect to the variable $\vec x$, given by $\mathcal{B}_i(\vec x)\equiv i\epsilon_{ijk} \braket{\partial_{j}\psi|\partial_{k}\psi}$.
For the $i$th component we get
\bean
\mathcal B_i\left(\vec x \right) = i\epsilon_{ijk} \braket{\partial_{l}\psi|\partial_{m}\psi} =i\epsilon_{ijk}\braket{\partial_j\psi(\vec x'(\vec x))|\partial_k\psi(\vec x'(\vec x))}=i\epsilon_{ijk}(\partial_j x'_l) (\partial_k x'_m)\braket{\partial'_{l}\psi|\partial'_{m}\psi},
\eean
where we used the chain rule.
The partial derivative $\partial_j x'_l=J_{lj}$ is an element of the Jacobian matrix.
This shows that the Berry curvature transforms as a 2-form.
Multiplying with the Jacobian from the left yields
\bean
J_{ni}\mathcal{B}_i\left(\vec x \right) =i\epsilon_{ijk}J_{ni}J_{lj}J_{mk}\braket{\partial'_{l}\psi|\partial'_{m}\psi}=i(\det \hat{\vec J}) \epsilon_{nlm}\braket{\partial'_{l}\psi|\partial'_{m}\psi}=(\det \hat{\vec J})\mathcal{B}'_n\left(\vec x'\left(\vec x\right) \right),
\eean
where the Berry curvature with the transformed argument $\mathbfcal{B}'(\vec x')$ appeared.
From this we write the transformation rule specific to 3-dimensional parameter space:
\bean\label{eq:berrytr1}
\mathbfcal{B}(\vec x)=(\det \hat{\vec J})\hat{\vec J}^{-1}\mathbfcal{B}'(\vec x'(\vec x)).
\eean
For a linear transformation $\vec x'(\vec x)=\hat{\vec J}\vec x$ the Jacobian is the coefficient matrix.

\subsection{Berry curvature in the parameter space of the ellipsoid}
Now we can use the transformation of the Berry curvature derived in Eq.~\eqref{eq:berrytr1} to the Berry curvature in Eq.~\eqref{eq:isoBerry} with the transformation introduced in Eq.~\eqref{eq:Btransform}
\bean
\mathbfcal{B}(\vec B)=(\det \hat{\vec J})\hat{\vec J}^{-1}\mathbfcal{B}'(\vec B'(\vec B)).
\eean
Here $\hat{\vec J}=\hat{\vec R}\hat{\vec g}_{\text R}^{\text T}$. Outside the ellipsoid we get
\bean\label{eq:Bout}
\mathbfcal{B}(\vec B_{S+})=\det\left(\hat{\vec R}\hat{\vec g}_{\text R}^{\text T}\right)\left(\hat{\vec R}\hat{\vec g}_{\text R}^{\text T}\right)^{-1}\frac{\hat{\vec R}\hat{\vec g}_{\text R}^{\text T}\vec B_S}{|\hat{\vec R}\hat{\vec g}_{\text R}^{\text T}\vec B_S|^3}=\det\left(\hat{\vec g}_{\text R}\right)\frac{\vec B_S}{|\hat{\vec g}_{\text R}^{\text T}\vec B_S|^3},
\eean
inside we get
\bean\label{eq:Bin}
\mathbfcal{B}(\vec B_{S_-})=0.
\eean

From Eq.~\eqref{eq:bnorm} we get the equation for the degeneracy ellipsoid $\vec B_S=B_S\vec b$
\bean\label{eq:bnorm2}
|\hat{\vec g}_\text{R}^{\text T}\vec B_S|=B_S|\hat{\vec g}_\text{R}^{\text T}\vec b|=B_S g_\text{R}=\frac{1}{2}\left(1+\frac{1}{a}\right).
\eean
To get an expression for the surface normal of the ellipsoid, we consider the scalar field
\bean\label{eq:ellipsoidnormalvector1}
f(\vec B)=\left(\hat{\vec g}_\text{R}^{\text T}\vec B\right)^2
\eean
that is constant on the degeneracy ellipsoid. Hence the gradient
\bean\label{eq:ellipsoidnormalvector2}
\vec\nabla f=2\hat{\vec g}_\text{R}\hat{\vec g}_\text{R}^{\text T}\vec B
\eean
is proportional to the normal vector of the surface.
Thus, the normal vector for the degeneracy ellipsoid at $\vec B_S$ reads 
\bean\label{eq:ellipsoidnormalvector3}
\vec n(\vec B_S)=\frac{\hat{\vec g}_\text{R}\hat{\vec g}_\text{R}^{\text T}\vec B_S}{|\hat{\vec g}_\text{R}\hat{\vec g}_\text{R}^{\text T}\vec B_S|}.
\eean
The surface topological charge density is proportional to the jump of the normal component of the Berry curvature
\bean\label{eq:surfdens}
\sigma(\vec B_S)=\frac{1}{2\pi}\left[\mathbfcal{B}(\vec B_{S+})-\mathbfcal{B}(\vec B_{S-})\right]\cdot\vec n(\vec B_S)=\frac{\det\hat{\vec g}_\text{R}}{2\pi|\hat{\vec g}_\text{R}^{\text T}\vec B_S|\cdot|\hat{\vec g}_\text{R}\hat{\vec g}_\text{R}^{\text T}\vec B_S|}=\frac{a\det\hat{\vec g}_\text{R}}{\pi(a+1)|\hat{\vec g}_\text{R}\hat{\vec g}_\text{R}^{\text T}\vec B_S|},
\eean
where in the last step Eq.~\eqref{eq:bnorm2} was used.

\end{widetext}

\end{document}